\documentclass[utf8]{frontiersSCNS} % for Science, Engineering and Humanities and Social Sciences articles
\usepackage{url,hyperref,lineno,microtype,subcaption}
\usepackage[onehalfspacing]{setspace}
\usepackage{amsmath}
\usepackage{amsthm}
\usepackage{multirow}
\usepackage{bm}
\usepackage{color} 

\newcommand\varmp{\mathbin{\vcenter{\hbox{%
  \oalign{\hfil$\scriptstyle-$\hfil\cr
          \noalign{\kern-.3ex}
          $\scriptscriptstyle({+})$\cr}%
}}}}

\def\keyFont{\fontsize{8}{11}\helveticabold }
\def\firstAuthorLast{Sawicki {et~al.}} %use et al only if is more than 1 author
\def\Authors{Jakub Sawicki\,$^{1}$, Rico Berner\,$^{2,3}$, Thomas L{\"o}ser\,$^{4}$ and Eckehard Sch{\"o}ll\,$^{1,3,5,*}$}
\def\Address{$^{1}$Potsdam Institute for Climate Impact Research, Telegrafenberg A 31, 14473 Potsdam, Germany \\
$^{2}$Institut f{\"u}r Mathematik, Technische Universit{\"a}t Berlin, Stra{\ss}e des 17. Juni 136, 10623 Berlin, Germany \\
$^{3}$Institut f{\"u}r Theoretische Physik, Technische Universit{\"a}t Berlin, Hardenbergstra{\ss}e 36, 10623 Berlin, Germany \\
$^{4}$Institut LOESER, Wettiner Stra{\ss}e 6, 04105 Leipzig, Germany\\
$^{5}$Bernstein Center for Computational Neuroscience Berlin, Humboldt-Universit{\"a}t, Philippstra{\ss}e 13, 10115 Berlin, Germany%\\
}
\def\corrAuthor{Eckehard Sch{\"o}ll}
\def\corrEmail{schoell@physik.tu-berlin.de}

\begin{document}
\onecolumn
\firstpage{1}

\title[Modelling tumor disease and sepsis]{Modeling tumor disease and sepsis by networks of adaptively coupled phase oscillators} 

\author[\firstAuthorLast ]{\Authors} %This field will be automatically populated
\address{} %This field will be automatically populated
\correspondance{} %This field will be automatically populated

\extraAuth{}% If there are more than 1 corresponding author, comment this line and uncomment the next one.
%\extraAuth{corresponding Author2 \\ Laboratory X2, Institute X2, Department X2, Organization X2, Street X2, City X2 , State XX2 (only USA, Canada and Australia), Zip Code2, X2 Country X2, email2@uni2.edu}

\maketitle

\begin{abstract}

In this study, we provide a dynamical systems perspective to the modelling of pathological states induced by tumors or infection. A unified disease model is established using the innate immune system as the reference point. We propose a two-layer network model for carcinogenesis and sepsis based upon the interaction of parenchymal cells and immune cells via cytokines, and the co-evolutionary dynamics of parenchymal, immune cells, and cytokines. Our aim is to show that the complex cellular cooperation between parenchyma and stroma (immune layer) in the physiological and pathological case can be qualitatively and functionally described by a simple paradigmatic model of phase oscillators. By this, we explain carcinogenesis, tumor progression, and sepsis by destabilization of the healthy homeostatic state (frequency synchronized), and emergence of a pathological state (desynchronized or multifrequency cluster). The coupled dynamics of parenchymal cells (metabolism) and nonspecific immune cells (reaction of innate immune system) are represented by nodes of a duplex layer. The cytokine interaction is modeled by adaptive coupling weights between the nodes representing the immune cells (with fast adaptation time scale) and the parenchymal cells (slow adaptation time scale) and between the pairs of parenchymal and immune cells in the duplex network (fixed bidirectional coupling). Thereby, carcinogenesis, organ dysfunction in sepsis, and recurrence risk can be described in a correct functional context.

\tiny
 \keyFont{\section{Keywords:} adaptive networks, cluster synchronization, coupled oscillators, pattern formation, sepsis, tumor disease, cytokine activity} %All article types: you may provide up to 8 keywords; at least 5 are mandatory.
\end{abstract}

\section{Introduction}\label{sec:intro}

Tumors and sepsis are diseases of different genesis. They have very different time scales and the therapies are completely different. But the age incidence, the risk factors, the outcome and the temporal recurrence behavior are similar. This justifies the attempt to describe tumor disease and sepsis with a unified disease model. A shift of the paradigm is proposed by choosing the nonspecific innate immune system as the reference point for both diseases~\cite{LOE20}. The innate immune system interacts with mutant cells and pathogens. Its role is to maintain the integrity of the organism by actively eliminating foreign organisms, degrading the organism's own damaged cells, and activating and coordinating wound healing. It can ward off about 99\% of all infections\cite{BOM18}; it arose with the beginning of multicellular life, and has grown and protected life to this day~\cite{MAL12,RIC19}. It is ubiquitously present in the organism, is responsive without significant dead time, and is based on dynamically balanced activator-inhibitor mechanisms. Its regulation is essentially decentrally organized. The innate immune system includes humoral and cellular components, the endothelium, and the tissue stroma. With the multitude of components interacting in the innate immune system, it is organized in a complex way and has a broad response spectrum.

The innate immune system is not completely resistant to disturbances; it can be brought to dysregulation. Tumor disease and sepsis are two prominent examples of this behavior. Although tumor disease and sepsis correspond to different medical conditions, after the initial stages a relatively uniform course develops\cite{WEI14d,SIN16b}. The variability of tumor disease results from the initial genetic state of the tumor cells, their subsequent mutations, epithelial-mesenchymal transition, and interactions with the innate immune system~\cite{LON11}. Recently, in case of cancer the crosstalk of tumor cells with immune cells has been investigated in more detail~\cite{ZHA21d}. In sepsis, the innate immune system is activated by infection and the clinical course is determined by the individual patient's initial condition. In both cases, the innate immune system can be brought to dysregulation and develop its own clinical picture such as cachexia, coagulation disorders, and organ failure. In the case of tumor disease, this is compounded by space-occupying lesions, tissue invasion and destruction through the nonphysiological production of proliferation factors, cytokines and chemokines. Possibly, not only the cytokine concentration, the cytokine mix but also the gradient of their increase is responsible for the extent of dysregulation of the innate immune system and thus for the disease consequences~\cite{ALT19,MOR13a}. 

The unified disease model with the \textit{innate immune system} as reference point is the basis for our modeling approach in terms of nonlinear dynamics of complex networks. Note that this is not a biochemical or genetic or cellular or tissue model of carcogenesis, as has been reviewed elsewhere~\cite{VIN10}, but it rather describes tumor and sepsis and recurrence risk in a dynamical functional context. Complex networks are an ubiquitous paradigm in nature and technology, with a wide field of applications ranging from physics, chemistry, biology, neuroscience, to engineering and socio-economic systems. Of particular interest are adaptive networks, where the connectivity changes in time, for instance, in chemical or biochemical systems~\cite{JAI01}, where the reaction rates adapt dynamically depending on the variables of the system, or in neuronal synaptic plasticity~\cite{MAR97a,ABB00,MEI09a,LUE16}, in epidemics~\cite{GRO06b}, and in biological or social systems~\cite{GRO08a}. Another focus of recent research in network science are multilayer networks, which are systems interconnected through different types of links \cite{BOC14,DE13,DE15,KIV14}. A special case of multilayer networks are multiplex topologies, where each layer contains the same set of nodes, and only pairwise connections between corresponding nodes from neighboring layers exist~\cite{ZHA15a,MAK16,LEY17a,AND17,SAW18c,OME19,RYB19,NIK19,DRA20,BER21,SAW21,SHE21}.

Synchronization is an important feature of the dynamics in networks of coupled nonlinear oscillators~\cite{PIK01,STR01a,ALB02a,NEW03,BOC18,BER21c}. Various synchronization patterns are known such as cluster synchronization where the network splits into groups of synchronous elements~\cite{DAH12}, or partial synchronization patterns such as chimera states where the system splits into coexisting domains of coherent (synchronized) and incoherent (desynchronized) states~\cite{KUR02a,ABR04,PAN15,SAW20,SCH20,SCH20b}. These patterns were also explored in adaptive networks~\cite{SEL02,AOK09,TIM14,KAS17,BER19,BER20b,BER20a}, and in particular in adaptive two-layer networks of phase oscillators~\cite{KAS18,BER20}. Moreover, the role of synchronization is an important aspect in the field of network physiology, where multi-component physiological systems continuously interact in an integrated network to coordinate their functions~\cite{BAS12b,IVA14,BAR15b,MOO16,LIN16d}. There are empirical studies dealing with the structural organization and functional complexity of human organism which demonstrate phase-synchronization as well as phase transitions~\cite{XU06a,CHE06b,IVA09,BAR12e} between different modes of synchronization in real physiological systems. In case of complex diseases, the progression from a healthy to sick state can be abrupt and may cause a critical transition~\cite{CHE12a,LIU12b,LIU13a,LIU13b}.

In this article, we propose a two-layer network model for carcinogenesis and sepsis based upon the interaction of parenchymal cells and immune cells via cytokines and the co-evolutionary dynamics of parenchymal, immune cells, and cytokines. Parenchyma is the bulk of {\em functional} substance in an organ, in contrast to the stroma, which refers to the {\em structural} tissue of organs or structures. In many organs the parenchyma consists of epithelial cells. We stress that our model is not a detailed model of organs or of biochemical processes but a functional model of dynamic interactions. Thus, cytokines are not modeled in terms of concentrations but rather in terms of the cytokine-mediated information flow between nodes within each layer and between the parenchymal and the immune layer, describing the cytokine activity. In the following, we refer to the stroma as the immune layer. The article is organized as follows: In Section~\ref{sec:physiology}, we provide a brief overview of the physiology of tumor disease and sepsis. The functional network model is introduced in Sec.\,\ref{sec:model}. Here, we discuss all variables, parameters, and their physiological meaning. Moreover, the methods and measures used for the subsequent numerical analysis are introduced and highlighted in their physiological context. In Section~\ref{sec:tumor}, we discuss various dynamical scenarios of the model simulations of tumor disease that are observed in the presence of pathological cells. Section~\ref{sec:sepsis} presents analogous computer simulations for sepsis. The results are summarized in Sec.\,\ref{sec:conclusion}.

%-------------------------
% Physiological description
%-------------------------
\section{Physiological description}\label{sec:physiology}

\subsection{Initial situation}
The function of the innate immune system is ensured in a site-independent manner by the coordinated activity of humoral and cellular components. Coordination also occurs via cytokines. With new cytokine sources building up over individual lifetimes, the stability of regulatory behavior in the innate immune system changes. Contributing systemic factors include inflammaging~\cite{FRA14,CAL17a}, activation of the coagulation system~\cite{FRA14,TRA99a}, increase of body fat~\cite{FAS04,GAI07,CHO14d}, lack of exercise~\cite{ELI17,FUL18}, age-related normal fibrosis~\cite{BEN71,NEM71}, chronic inflammation as local factors~\cite{VIR78}, or smoking via initiation of local hypoxia \cite{BRU18b}. The innate immune system exhibits pro-inflammatory activation. The number of individual variables suggest a wide range of activation levels, which still depend on cytokine gene polymorphisms, among other factors. Physiological cytokine production and additional pro-inflammatory cytokine activity (pattern, concentration, emitters involved, gradient of increase) determine the activation level of the innate immune system.\\

\subsection{Tumor disease}
In tumor disease, mutant cells interact with macrophages of the innate immune system localized in the stroma. Mutated cells are genetically altered and genetically unstable parenchymal cells. They are a new cell entity with altered regulatory behavior. In interaction with the normal stroma (immune layer), e.g. in young healthy individuals, mutant cells do not find survival conditions. They persist silently or trigger apoptosis via tissue surveillance~\cite{LOE18a}. Mutant cells in an area pre-damaged by chronic inflammation receive cytokines from the activated stroma, undergo an epithelial-mesenchymal transition~\cite{LAM14a,CHO16c}, can proliferate, disseminate, and reprogram normal macrophages into tumor-associated macrophages~\cite{WU09a,KAR17a,MAN17}. They help to ensure oxygen and nutrient supply to the tumor. Tumor cells, tumor-associated macrophages and normal stromal cells communicate via cytokines. The communication strength depends on tumor cell genetics and stroma activation. It increases with the tumor cell mass and the mass of specific tumor stroma. A mutually activating circular process starts between the tumor cells and the innate immune system. A specific cytokine microclimate is formed around the developing tumor, the size of which grows with the increase in tumor mass. Cytokines enter the lymph nodes with the lymph and enable the tumor cells that have floated there to proliferate. Lymphogenic metastasis begins, continuing downstream in additional lymph nodes according to the same pattern. Organ-specific metastatic patterns are formed~\cite{SCH48}. The circulating cytokine concentration increases and triggers the general symptoms of tumor disease such as cachexia~\cite{ARE15} and finally the lethal coagulopathy, lung failure, organ failure or lethal space-occupying lesions and tissue destruction. The postulate of Lewis Thomas~\cite{THO72} also applies to tumor disease, according to which the interactions with the organism triggered by the tumor cells are responsible for the disease, i.e., tumor growth, enabling malignant cells to become invasive and destructive, recurrence, and finally induce the lethal general symptoms. Hematogenous metastasis may start when the disseminated tumor cells are adequately supplied with growth factors (cytokines) via further activation of the innate immune system.\\

\subsection{Sepsis}
The initial conditions for the transition of an infection into sepsis are analogous to the survival conditions of mutated cells in the tissue, a pre-activated innate immune system. Sepsis is triggered when the innate immune system no longer succeeds in locally fixing invading pathogens, but allows them to enter the organism. There they interact with the already pre-activated innate immune system. The immune system is activated systemically. Antibiosis or surgical sanitation can reduce or eliminate the pathogen load. Depending on the type of reaction, the pro-inflammatory response is stopped or it continues to escalate in varying degrees~\cite{SEY19}, for which also cytokine gene polymorphisms are responsible~\cite{HOT16,MAJ01,THO20}. Pro-inflammatory cytokines act on endothelial cells and the coagulation system. Microthrombi amplify inflammation through secretion of growth factors and cytokines \cite{HOT16}. The endothelial permeabilization barrier is opened, fluid retention in tissues occurs, blood pressure drops and must be stabilized with fluid administration \cite{BRU18b}. Prolonged oxygen diffusion pathways and the resulting tissue hypoxia trigger cytokine release causing a "second hit", resulting in metabolic changes in the parenchyma~\cite{BOM18}. Organ failure, particularly of the kidneys, lungs, and liver is imminent. Endogenous molecules with immunogenic effects, especially mitochondrial DNA \cite{FRA18a}, released e.g. after trauma, burns, pancreatitis or surgery, can have the same effect as exogenous pathogens.\\

\subsection{Relapse}
Tumors and sepsis have a recurrence behavior, the frequency of which correlates with the stage of the primary disease. In both diseases, the activated innate immune system is an initial prerequisite for the primary disease, which is brought to dysregulation in the course of the disease. After tumor removal and elimination of infection, the activation status reduces but remains at least at the level before disease onset. Thus, cytokines are produced for remaining tumor cells as a result of pro-inflammatory activation~\cite{ARE15,GAS82,EIC04,LIP16}, which promote their proliferation and allow the organization of the tumor stroma. The process kinetics is determined by the proliferation rate of tumor cells, the availability of cytokines, and by the ability of tumor cells to organize their own stroma with connection to the blood supply. After clearance of the systemic infection, the innate immune system has an activation status at least equivalent to that before septic shock. The infection itself appears to have triggered irreversible elements of trained immunity~\cite{BOM18}. Surviving patients after septic shock thus die in 50\% within the first two post-sepsis years from persistent and secondary nosocomial infections, tumors, or cardiac failure~\cite{PRE16a}.\\

%-------------------------
% Model
%-------------------------
%\vspace{0.5cm}
\section{Model}\label{sec:model}
\subsection{Schematic model for tumor disease and sepsis}
\begin{figure}[htbp]
	\centering
	\includegraphics[width = 1.0\textwidth]{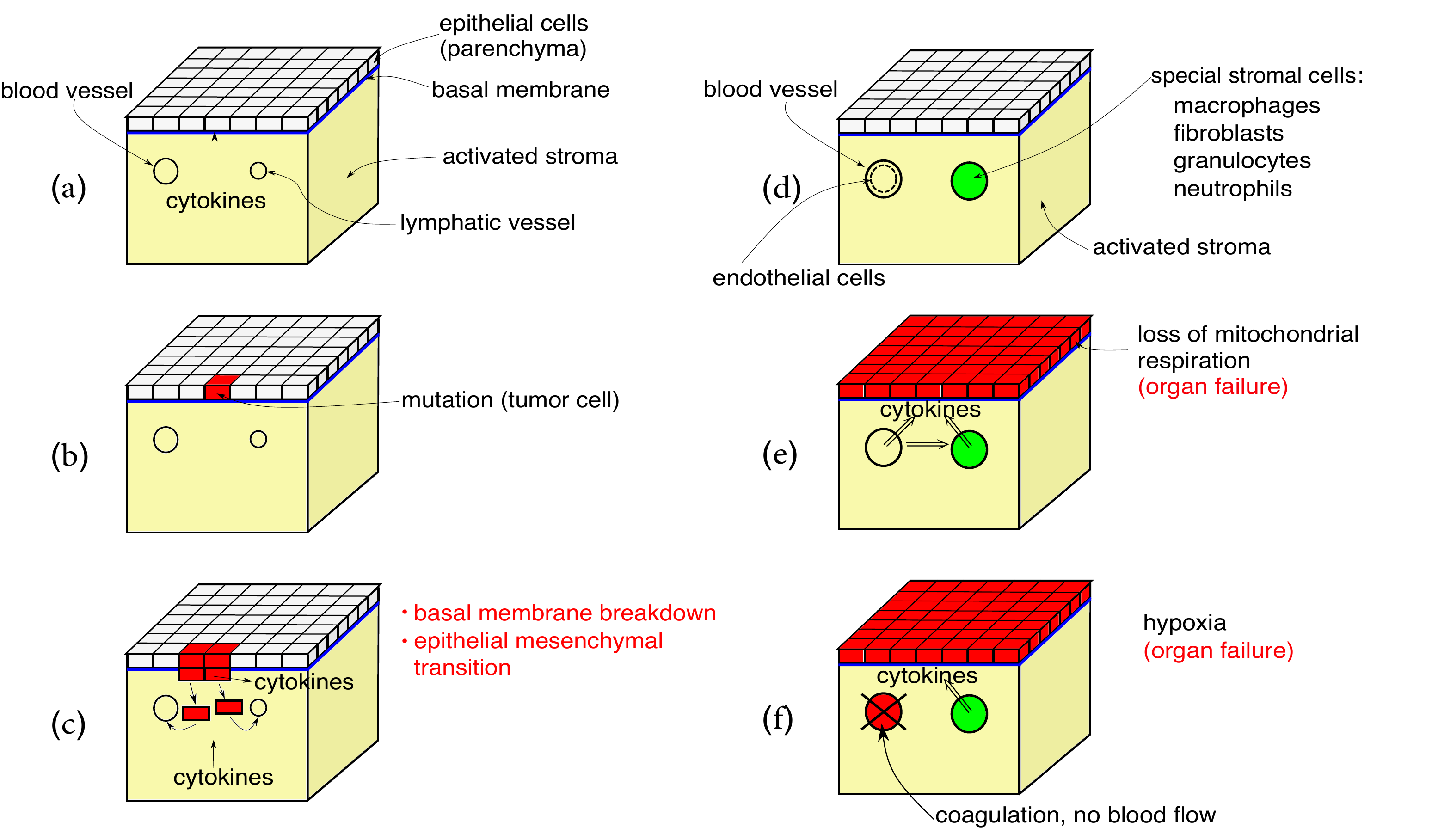}
	\caption{Scheme of a tissue element showing the initial progression of tumor disease (a)-(c), the initial configuration for sepsis (d) and the final configurations found for both tumor disease and sepsis (e),(f).
	}
	\label{Fig1}
\end{figure}

Organic tissue consisting of parenchymal cells and immune cells is shown schematically in Fig.\,\ref{Fig1}. We depict the initial and final configurations for tumor disease and sepsis on a tissue element. The tissue element consists of the epithelial parenchyma, the basal membrane and the stroma. The parenchyma is the organ-specific functional layer. The basal membrane separates the parenchyma and stroma and is made of collagen of type IV. In the stroma, blood supply, lymphatic drainage, and immune response occur. The stroma consists of an extracellular matrix and embedded cells that do not form a solid association. The extracellular matrix is structurally composed of collagen, glycoproteins, proteoglycans, and water. Cells in the stroma are resident fibroblasts and fat cells, and mobile cells (macrophages, mast cells, granulocytes, and plasma cells).

The initial process of carcinogenesis is shown in Fig.\,\ref{Fig1}(a)-(c). In Fig.\,\ref{Fig1}(a), the parenchyma is normal, the stroma is inflammatorily activated. In Fig.\,\ref{Fig1}(b), a mutation of a tumor cell has occurred in the parenchyma, the inflammatory activation of the stroma continues. Finally in Fig.\,\ref{Fig1}(c), the mutant cells proliferate, break through the basal membrane and migrate into the blood and lymphatic vessels, and cytokines are emitted.

Figure~\ref{Fig1}(d)-(f) shows the systemic effects of the final process of tumor disease and sepsis: Figure \ref{Fig1}(d) depicts a tissue element that is not directly affected by tumor, metastasis, or primary inflammation. The stroma exhibits inflammatory activation. In Fig.\,\ref{Fig1}(e), a systemic cytokine storm is occurring. Cytokine production by the stromal cells is additionally stimulated. The parenchyma responds to the primary and secondary cytokine storm by uncoupling mitochondrial respiration and switching to aerobic glycolysis. The energy supply is no longer sufficient for organ-specific cellular functions and the organ fails. Figure~\ref{Fig1}(f) shows a variant of Fig.\,\ref{Fig1}(e). The cytokine storm activates the endothelium, blood coagulation is activated, oxygen transport breaks down, and hypoxia arises. The stromal cells respond with cytokine release. Both together, the hypoxia and the cytokine release, lead to the collapse of organ-specific cell functions in the parenchyma and the organ fails.\\
\\

\subsection{Dynamical two-layer network model}
The unified disease model with the reference point given by the \textit{innate immune system} is the basis for our model, which includes disease-specific initial conditions, mutant cells for tumor disease and infection-driven cytokine dysregulation. A volume element of tissue consisting of parenchyma, basal membrane and stroma is used as a model for tumor disease and sepsis, describing the functional interactions between parenchyma (organ tissue) and stroma (immune layer). We represent the network layer of parenchymal cells (superscript 1) by $N$ phase oscillators $\phi_i^{1}$, $i=1, \ldots, N$, with partly fixed and partly adaptive coupling weights~\cite{JUN95}, and the network layer of immune cells (superscript 2) by $N$ adaptively coupled phase oscillators $\phi_i^{2}$. The communication through cytokines which mediate the interaction between the parenchymal cells is modeled by the coupling weights $\kappa_{ij}^{1}$, and those between the immune cells by coupling weights $\kappa_{ij}^{2}$.\\ 

The use of phase oscillators for the functional modeling of the interacting parenchymal cells and immune cells is motivated by the fact that phase oscillator networks are a paradigmatic model for collective coherent and incoherent dynamics. As discussed in detail in Sect. 3.3, healthy cells and tumor cells differ by their metabolic activity, i.e., tumor cells are less energy-efficient and thus have a faster cellular metabolism, which is reflected in our phase oscillator model by a higher frequency. Thus in the healthy homeostatic equilibrium state all parenchymal cells have the same lower frequency, while the pathological state splits into two clusters with different frequencies, healthy and unhealthy, i.e., a multifrequency cluster. The healthy state is assumed to be characterized by regular periodic, fully synchronized dynamics of the phase oscillators, i.e., all cells show the same collective frequency of cellular metabolism. The pathological state is described by multifrequency clusters with different frequencies, i.e., the pathological cells in the parenchyma attain a higher frequency, while the healthy cells are still frequency-synchronized with the ``healthy'' frequency. This loss of synchrony reflects the fact that in case of tumor disease, the malignant mutation basically leads to a loss of performance of the parenchymal cells. As a consequence, they are no longer fully coordinated, which leads to a loss of proliferation and apoptosis control~\cite{LON11,WEI14d}. Such a pathological condition leads also to an alteration of the metabolic activity of the immune cells~\cite{COU02,HEE15a,CHO16c,POR16b,RAZ18,GRE19}. In case of sepsis, i.e., a systemic inflammation, cells belonging to the innate immune system produce cytokines in an unregulated way affecting the parenchyma or microcirculation leading to organ failure~\cite{SIN16b,MAT18c} which is also associated in our model with the loss of synchrony.

A general multiplex network with two layers each consisting of $N$ identical adaptively coupled phase oscillators is described by
\begin{align}
\label{eq:DGL_somatic}
	\dot{\phi}_{i}^{1} &=\omega_i^{1}-\frac{1}{N}\sum_{j=1}^{N}(a_{ij}^1+\kappa_{ij}^1)\sin(\phi_{i}^1-\phi_{j}^1+\alpha) %\nonumber \\[-.15cm]
	-\sigma \sin(\phi_{i}^1-\phi_{i}^2), \\
		\dot{\kappa}_{ij}^1&=-\epsilon^1 \left (\kappa_{ij}^1+\sin(\phi_{i}^1-\phi_{j}^1-\beta)\right), \nonumber
\end{align}
\begin{align}
\label{eq:DGL_immune}
	\dot{\phi}_{i}^{2} &=\omega^{2}-\frac{1}{N}\sum_{j=1}^{N}\kappa_{ij}^2\sin(\phi_{i}^2-\phi_{j}^2+\alpha) %\nonumber \\[-.15cm]
	-\sigma \sin(\phi_{i}^2-\phi_{i}^1), \\
	\dot{\kappa}_{ij}^2&=-\epsilon^2 \left (\kappa_{ij}^2+\sin(\phi_{i}^2-\phi_{j}^2-\beta)\right), \nonumber
\end{align}
where $\phi_i^{\mu}\in [0,2\pi)$ represents the phase of the $i$\textsuperscript{th} oscillator ($i=1,\dots,N$) in the $\mu$\textsuperscript{th} layer ($\mu=1,2$), $\omega_i^1\equiv \omega_i$ are the natural oscillator frequencies of the parenchymal cells which are distributed according to a probability distribution $\rho(\omega^1)= (1-r)\delta(\omega^1-\omega^h) + r\delta(\omega^1-\omega^p)$ where $r$ is the fraction of pathological parenchymal cells relative to the number of all parenchymal cells $N$, $\delta$ is the Dirac delta function, and $\omega^p$ and $\omega^h$ are the natural frequencies of pathological and healthy parenchymal cells, respectively. The value of $\omega^2 \equiv \omega$ is the natural frequency of the immune cells. The interaction between the oscillators within each layer is determined by the intralayer connectivity weights $a_{ij}^1\in[0,1]$ (fixed interaction within an organ) and $\kappa_{ij}^{\mu}\in[-1,1]$ (adaptive interaction mediated by cytokines), where the parenchymal layer has both fixed and adaptive couplings, while the immune layer has only adaptive coupling. Between the layers the interlayer coupling weights $\sigma \ge 0$ are fixed and symmetric for both directions of interaction. The parameter $\alpha$ is a phase lag of the interaction modeling a time-delay \cite{SAK86,ASL18a}. The adaptation rates $0<\epsilon^\mu \ll 1$ separate the time scales of the slow dynamics of the coupling weights and the fast dynamics of the oscillatory system. The adaptation rate of the parenchymal layer $\epsilon^1$ is assumed to be slow compared to the adaptation rate of the immune layer $\epsilon^2$, i.e., $\epsilon^1 \ll \epsilon^2$ to account for the faster reaction of the immune cells~\cite{MOR13a,ALT19}. Thus we have two classes of adaptive coupling weights modeling two different cytokine mechanisms on two different timescales. As a consequence of choosing two significantly different values for $\epsilon^1$ and $\epsilon^2$, we obtain a system with multiple times scale dynamics, i.e., "slow-fast-faster" dynamics ($\epsilon^1\ll\epsilon^2\ll 1$)~\cite{KUE15}.\\

The phase lag parameter $\beta$ of the adaptation function $\sin(\phi^\mu_i-\phi^\mu_j-\beta)$, also called plasticity rule in the neuroscience terminology~\cite{AOK09}, describes different adaptation rules that may occur. For instance, for $\beta=\varmp\pi/2$, a symmetric rule~\cite{HOP96,SEL02,AOK15,ROE19a} is obtained where the coupling $\kappa_{ij}$ decreases (increases) between any two systems with close-by phases. If $\beta=0$, the link $\kappa_{ij}$ will be strengthened if the $i$\textsuperscript{th} oscillator is advancing the $j$\textsuperscript{th}. Such a causal relationship is typical for spike-timing dependent plasticity in neuroscience \cite{CAP08a,MAI07,LUE16,POP13}. The matrix elements $a_{ij}^1\in\{0,1\}$ of the adjacency matrix $A$ in the parenchymal layer are chosen as $a_{ij}^1=1$ if $i\ne j$ (global coupling). For normalization of the coupling terms the coupling sums in Eqs.\,(\ref{eq:DGL_somatic}) and (\ref{eq:DGL_immune}) are multiplied by the normalization factor $\frac{1}{N}$.\\
\\

\subsection{Methodology and Measures}
Heterogeneous dynamics, e.g., multifrequency clusters and tumor growth, may arise through the dynamic interaction of parenchymal cells, immune cells, and localized cytokine activity in a self-organized self-adaptive manner, even if the system parameters in the layers are chosen uniformly, i.e., homogeneous~\cite{BER19a,BER20c}. Heterogeneity can enter the parameters of oscillator networks in various ways. Very prominent are heterogeneities in the natural frequencies of the oscillators or in the connectivity structure~\cite{ACE05}. Heterogeneous frequencies $\omega_{i}$ are used to model pathological parenchymal cells which occur by spontaneous random mutations. We assume that all healthy cells possess the same natural frequency $\omega_{i}=\omega^h$, and a fraction $r$ of randomly chosen pathological cells possess the natural frequency $\omega_{i}=\omega^p$. The characterization of tumor cells via their metabolic properties~\cite{LON11,WEI14d} has revealed a difference in the metabolic activity between tumor and healthy cells~\cite{WAR24}. 
Warburg demonstrated that cells after a malignant mutation obtain their energy via aerobic glycolysis, which provides only 4 mol ATP/mol glucose, in contrast to normal cells whose metabolism is based upon mitochondrial breathing yielding 36 mol ATP/mol glucose. Thus tumor cells have a metabolic efficiency of only about 10\% of healthy cells, and therefore they need more glucose. Other general differences in metabolic activity between tumor cells and healthy cells include proliferation rate, apoptosis rate, initiation of angiogenesis, tissue invasion or metastatic capacity. Here, quantitative differences exist between different organ tumors. Ultimately, any change in metabolic performance, regardless of which metabolic branch is affected (structural, energetic, cell division, or special metabolism), can be conceived as a frequency change. 

Using numerical simulations, we study whether the healthy state (frequency synchronized) is persistent also in the presence of a few pathological cells. Under certain conditions depending on various parameters (age, inflammaging, chronic inflammation, other basic diseases, obesity, smoking, lack of exercise), an unregulated cytokine expression and hence a possibly lethal tumor can occur. In these cases, the healthy (synchronized) state is not persistent anymore against the perturbation by the heterogeneity (tumor cells). For our study, the cytokine adaptivity parameter (which we call {\em age parameter}) $\beta$ and the fraction $r$ are considered as the main model parameters to account for the various system conditions. In the case of sepsis, we introduce a fixed initial perturbation of the cytokine activity in the immune layer representing a systemic immune response, while keeping the natural frequencies uniform. Similar to the tumor disease, we study the effect of this initial system perturbation on the emergence of the healthy state in dependence of the age parameter $\beta$. For our simulations we have used a Runge–Kutta method of order 4 with a fixed stepsize of $\Delta t= 0.05$ and simulation time of $T_s= 2000$ time units where we have discarded the first $1000$ time units to account for transient dynamics.

In order to quantitatively characterize the dynamic collective state of the two-layer network, we introduce several measures. First, we use the mean phase velocities of the oscillators $j$ in both layers $\mu=1,2$
\begin{align}\label{avFreq}
	\langle \dot{\phi}_j^\mu \rangle = \frac1T \int_{t}^{t+T}\dot{\phi}_j^\mu(t') \mathrm{d}t' = \frac{{\phi}_j^\mu(t+T)-{\phi}_j^\mu(t)}{T}
\end{align} 
with averaging time window $T$, and the spatially averaged mean phase velocity for each layer $\bar{\omega}^\mu=\frac{1}{N}\sum_{j=1}^{N} \langle \dot{\phi}_j^\mu \rangle$. 

Furthermore, for both layers $\mu=1,2$ we calculate the ensemble average $s^\mu$ (ensemble size $N_E$ with ensemble elements $E$) of the standard deviation $\sigma_\chi(\bar{\omega}^\mu)= \sqrt{\frac1N \sum_{j=1}^N (\langle \dot{\phi}_j^\mu \rangle-\bar{\omega}^\mu)^2}$ of the mean phase velocities 
\begin{equation}
	\label{eq:sd}
	s^\mu= \frac{1}{N_E}\sum_E\sigma_\chi(\bar{\omega}^\mu_E), 
\end{equation}
and the ensemble average of the corresponding normalized standard deviation $\frac{\sigma_\chi(\bar{\omega}^\mu_E)}{\bar{\omega}^\mu_E}$. If the latter quantities are non-zero, they indicate the formation of multifrequency clusters, where the respective layer splits into clusters with different frequencies, which is indicative of a pathological state.

In systems of adaptively coupled phase oscillators one often encounters frequency-synchronized dynamics. If the frequency is the same, the phases may still exhibit different behavior. They may either be all the same (complete in-phase synchronization) or they may be phase-locked such that each phase oscillator oscillates with the same frequency but a fixed, time-independent phase difference. A special case is a splay state, where the phase differences of all oscillators average out, for instance if the relative phase of the $j$-th oscillator is $2\pi j/N$, $j=1,\ldots,N$. In systems of the form Eqs.~(\ref{eq:DGL_somatic}) and (\ref{eq:DGL_immune}), it is possible to find in-phase synchronization and splay states, and they may be interpreted as different quality of synchronization~\cite{BER20,BER21e}. In our set-up a splay state is interpreted as a more vulnerable collective state (synchronization of parenchyma and immune layer) where small perturbations can quickly lead to partial or complete desynchronization.\\

\subsection{Physiological interpretation of model parameters}

\begin{table}
	\begin{center}
		\begin{tabular}{|c||c l l|} 
			\hline  
			& symbol & name & physiological meaning \\ [0.5ex] 
			\hline\hline
			dynamical & $\phi_i$ & phase & metabolic activity \\
			\cline{2-4}
			variable & $\kappa_{ij}$ & coupling weight & cytokine activity \\ %information potential, cytokine information transfer, cytokine mediated cell-cell interaction strength
			\hline \hline
			& $\alpha$ & phase lag & metabolic interaction delay\\
			\cline{2-4}
			& $\beta$ & plasticity rule & age, inflammaging, diseases, malignity (tumor disease) \\
			\cline{2-4}
			& $\omega$ & natural frequency & natural frequency of cellular metabolism\\ %Grundumsatz
			\cline{2-4}
			parameter & $\epsilon$ & time scale ratios & time scales of cytokine activity\\
			\cline{2-4}
			& $r$ & inhomogeneity ratio & fraction of tumor cells\\ 
			\cline{2-4}
			& $a_{ij}$ & connectivity & fixed parenchymal cell-cell interaction\\
			\cline{2-4}
			& $\sigma$ & interlayer coupling & interaction between immune \& parenchymal cells\\
			\hline \hline
			& $\langle \dot{\phi_i}\rangle$ & mean phase velocity & collective frequency of cellular metabolism\\
			\cline{2-4}
			measure & $s$ & standard deviation & pathogenicity (parenchymal layer),\\
			& & (see Eq.\eqref{eq:sd}) & activation (immune layer) \\
			\hline
		\end{tabular}
		\caption{\label{tab:parameter}Transcription table for the dynamical variables, parameters and measures of the model (superscripts referring to layers $\mu=1$ and $\mu=2$ omitted).}
	\end{center}
\end{table}

Table~\ref{tab:parameter} gives an overview of the dynamical variables, parameters and measures of the model. In the right column the physiological meaning of all quantities is given in a concise manner. In more detail, the quantities of the mathematical model which are listed in Table~\ref{tab:parameter} have the following physiological correspondence:

\renewcommand{\labelenumi}{(\roman{enumi})}
\begin{enumerate}
  \item Variable $\phi_i$ mimics the metabolic activity of a parenchymal or immune cell $i$ as a universal oscillatory phase variable. Here our aim is not to describe physiological processes on a detailed biochemical level, e.g., C-reactive protein (CRP) production after inflammatory activation, etc.
  \item Variable $\kappa_{ij}$ describes the cytokine activity mediating information flow between cells $j$ and $i$. It stands for the susceptibility of communications between cells via cytokines. %(genetically determined). 
Cytokines have pleiotropic properties, i.e., they can exert different, even contrary, effects depending on the initial situation and in combination with other cytokines. Therefore we model them by an adaptive coupling strength. Reaction strength of cytokines is due to individually different gene polymorphisms.
  \item Parameter $\alpha$ models the phase lag of the intralayer cell-to-cell coupling. Its origin is the time delay from the activation of a defined metabolic branch until the mediator release; within the inflammatory cascade, characteristic time delays exist for all reaction sub-steps starting from the triggering of the inflammatory reaction, for example, the reaction time of the individual liver cells starting from the input signal until the maximum blood concentration of interleukin IL-6 (2h) and CRP (12h) is reached.
  \item Parameter $\beta$ governs the adaptivity (or plasticity) rule of the cytokines. It mimicks a systemic sum parameter which may account for different influences such as physiological changes due to age in the extracellular matrix, inflammaging, systemic and local inflammatory baseline, adiposity, pre-existing illness, physical inactivity, nutritional influence, and others. In case of tumor disease, it can include the malignancy grade of tumor cells. For the sake of brevity, we call this parameter the age parameter.
  \item Parameter $\omega$ denotes the natural frequency of the basic metabolic activity of a single cell. For instance, it can stand for the minor CRP production in a normal healthy state.
  \item Parameters $\epsilon^1$ and $\epsilon^2$ denote the inverse relaxation times (half-life) of the cytokines in the parenchymal and immune layer, respectively.
  \item Parameter $r$ denotes the fraction of mutant (pathological) parenchymal cells in the tissue element, and is implemented in the model as the fraction of cells with a deviating (pathological) natural frequency $\omega^p$.
  \item Parameter $a_{ij}$ denotes the genetically fixed intercellular communication pathways between parenchymal cells, they are \textit{wired} by fixed cell-to-cell connections. In contrast, communication of parenchymal cells via cytokines and of cells within the immune layer runs in an open communication channel, which is controlled self-adaptively, after~\cite{SHA48}.
  \item Parameter $\sigma$ denotes the interlayer coupling strength between the parenchymal and immune cells. It can be due to a mass transfer, e.g. cytokine expression of macrophages in the immune layer, and signal transfer into the parenchyma: macrophages invade the extracellular matrix and penetrate the basal membrane to the parenchyma.
  \item The mean phase velocity $\langle \dot{\phi}_i\rangle$ is a measure which describes the collective frequency of the cellular metabolism of cell $i$ due to all interactions with other cells in both layers. It denotes the system performance of a single metabolic branch of the parenchymal or immune cells of a tissue element according to their activation state (normal healthy state or pro-inflammatory activation with increased CRP production).
	%$=$ basal metabolic activity of CRP concentration in blood $<5$mg/l, in pro-inflammatory activated state CRP production is ramped up to blood concentration within $6-12$h and can assume values of $\gg 100$mg/l)
  \item The ensemble-averaged standard deviation of the mean phase velocities $s^1$, $s^2$ is a measure which characterizes the inhomogeneity of metabolic activity within the parenchyma and the stroma (immune layer), respectively. It assumes non-zero values if the respective layer is not frequency-synchronized and splits into multifrequency clusters, e.g., a healthy cluster with one frequency and a pathological cluster with another frequency, and thus is a measure of pathogenicity in case of the parenchymal layer, or activation in case of the immune layer.
		
	%as a function of the number of mutant cells, e.g. for the production of tumor markers, cytokines, or cell cycle time.  
\end{enumerate}

%-------------------------
% Methods
%-------------------------
%\section{Methods}\label{sec:Methods}

%-------------------------
% Tumor disease
%-------------------------
\section{Tumor disease}\label{sec:tumor}

In this section, we present exemplary computer simulations of our model to demonstrate different dynamic scenarios which this model can produce already in its simplest form. The model has not been refined or optimized with respect to the parameters, but our concern here is to display simulations which can describe principally different evolutions and outcomes of tumor disease. We assume $N=200$, $\epsilon^1=0.03$, $\epsilon^2=0.3$, $a_{ij}^1=1$ (global coupling in the parenchymal layer), $\sigma=0.3$ (interlayer coupling), natural frequencies $\omega^h=\omega^2=0$ (co-rotating frame of healthy parenchymal cells), $\omega^p=1$ (pathological cells describing spontaneous mutation, chosen randomly for $R$ oscillators where $r=R/N$), and the coupling delay parameter $\alpha$ is chosen the same for both layers such that the frequencies of the healthy parenchymal and immune layers are similar. The parameters $\beta$ is used to model the influence of age, inflammation status, environment etc. and also the degree of malignancy of the tumor.

\begin{figure}[ht]
	\centering
	\includegraphics[width = .85\textwidth]{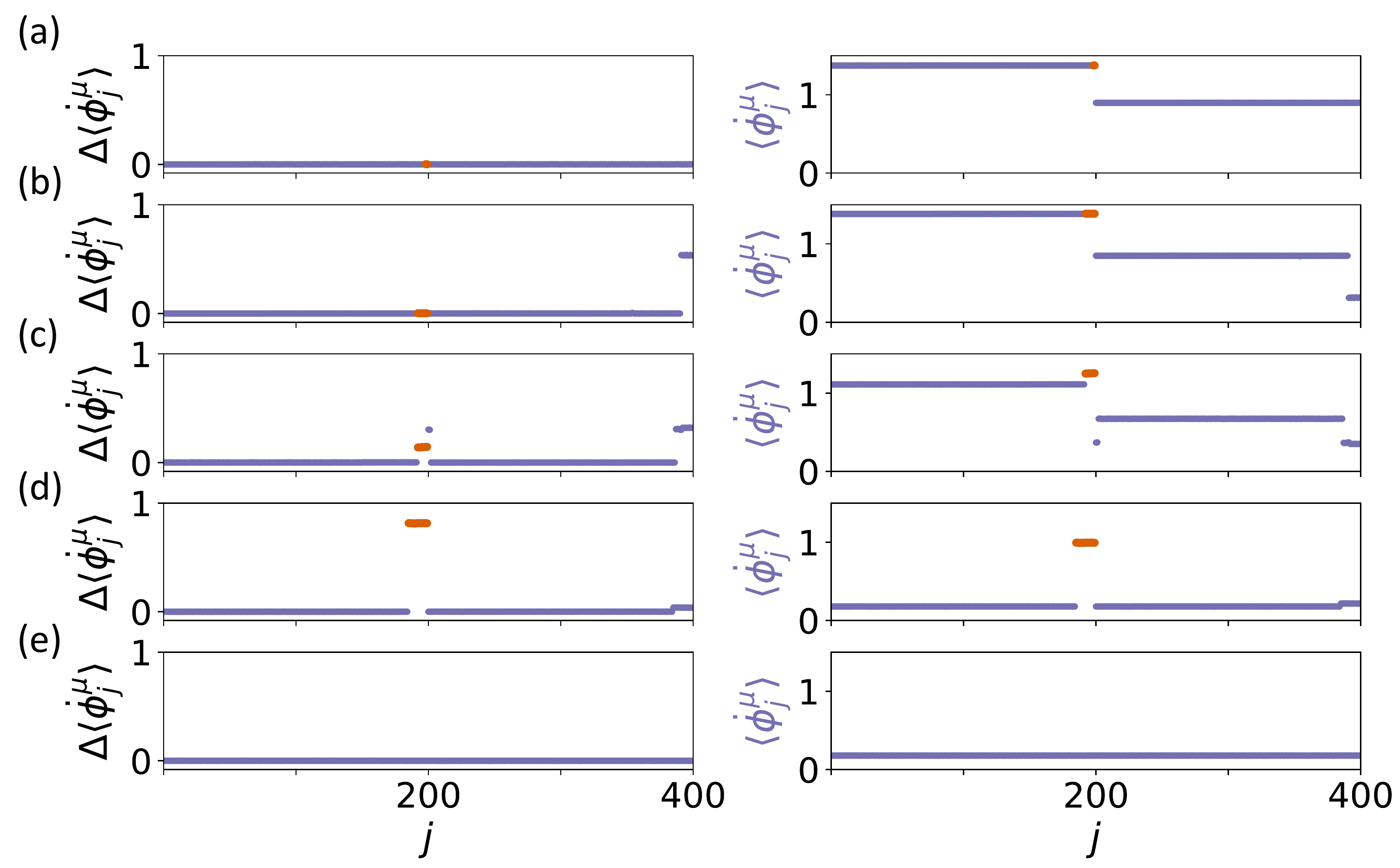}
	\caption{Dynamical scenarios for healthy states without clusters and pathological states with multifrequency clusters for different values of $\beta$ and $r$. %marked in Fig.\,\ref{Fig4}a,b. 
	The left panel shows the absolute differences of the mean phase velocities $\Delta\langle\dot{\phi}_j^\mu\rangle =|\langle \dot{\phi}_j^{1/2} \rangle-\langle \dot{\phi}_{100/300}^{1/2} \rangle|$, whereas the right panel shows mean phase velocities $\langle \dot{\phi}_j^\mu \rangle$. The parenchymal nodes are labeled $j=1,...,200$, and the immune nodes $j=201,...,400$ for $\beta=0.45\pi$ and $r=1\%$ (a), $\beta=0.55\pi$ and $r=4\%$ (b), $\beta=0.60\pi$ and $r=4\%$ (c), $\beta=0.66\pi$ and $r=7.5\%$ (d), and $\beta=0.66\pi$ and $r=0\%$ (e). Within each layer $\mu$ the nodes are sorted first by $\langle \dot{\phi}_j^1 \rangle$, then by $\phi_j^1$, respectively. Simulation parameters: $N=200$, $\alpha=-0.28\pi$, $a^1_{ij}=1$, $\epsilon^1=0.03$, $\epsilon^2=0.3$, $\sigma=0.3$, $\omega^h=\omega^2=0$, $\omega^p=1$. The simulation time is $2000$ time units. For the mean phase velocities the last $T=1000$ time units are taken for the temporal average, see Eq.\,\eqref{avFreq}. The mean phase velocities of nodes in layer $\mu=1$ with $\omega^p=1$ are marked red. (c), (d) show pathological states.
	}
	\label{Fig2}
\end{figure}

Our system exhibits various dynamic patterns resembling healthy and pathological states. The results are depicted in Fig.\,\ref{Fig2}, where the mean phase velocities $\langle \dot{\phi}_j^\mu \rangle$ are plotted in the right panel, and the absolute differences of the mean phase velocities with respect to a reference oscillator in the parenchyma $\langle \dot{\phi}_{100}^{1} \rangle$ and in the immune layer $\langle \dot{\phi}_{300}^{2} \rangle$, respectively, is plotted in the left panel. Fig.\,\ref{Fig2}(a) shows a healthy state with a low age parameter $\beta=0.45\pi$ and only two tumor cells ($r=1\%$) in the parenchymal layer. Although the two tumor cells in the parenchyma have different frequencies, the frequency-synchronized state with the healthy cells can be maintained due to the coupling. The coupled cooperative dynamics leads to a completely in-phase synchronized state in both layers. Note that the collective frequencies in the parenchymal layer and the immune layer are different (right panel) since the coupling terms are different. The same synchronized state of the parenchyma holds in Fig.\,\ref{Fig2}(b) for increased age parameter $\beta=0.55\pi$ and increased number of tumor cells ($r=4\%$), but first indications of pathological behavior are visible in the changed activity of the immune layer, showing lower frequencies of the associated immune cells coupled to the tumor cells (right panel, $j=393, \ldots, 400$). For higher age parameter $\beta=0.60\pi$ and the same number of tumor cells ($r=4\%$), Fig.\,\ref{Fig2}(c) shows a scenario of an emergent pathological state, i.e., a two-frequency cluster with higher frequency of the pathological cells (red), in the parenchymal layer. A corresponding cluster of lower frequencies is observed in the immune layer (Fig.\,\ref{Fig2}(c) as well. With increasing age parameter $\beta=0.66\pi$ and tumor size $r=7.5\%$, Fig.\,\ref{Fig2}(d) shows an even more pathological state, where the frequency of the split-off pathological cluster (red) in the parenchyma is distinctly more different from the healthy cluster (blue), while the immune layer is no longer supporting a two-cluster state but has adjusted in frequency to the parenchyma (right panel). This indicates an essential change in the dynamic state which is frequency synchronized close to zero, which indicates the absence of strong coupling contributions to the frequency. This frequency synchronization between parenchymal and immune layer is maintained if the large age parameter $\beta=0.66\pi$ is kept, but the number of tumor cells is chosen as zero (Fig.\,\ref{Fig2}(e)), and a healthy completely frequency-synchronized state is obtained. In the following, we will characterize this new dynamic state in more detail.

\begin{figure}[ht]
\centering
\includegraphics[width = .9\textwidth]{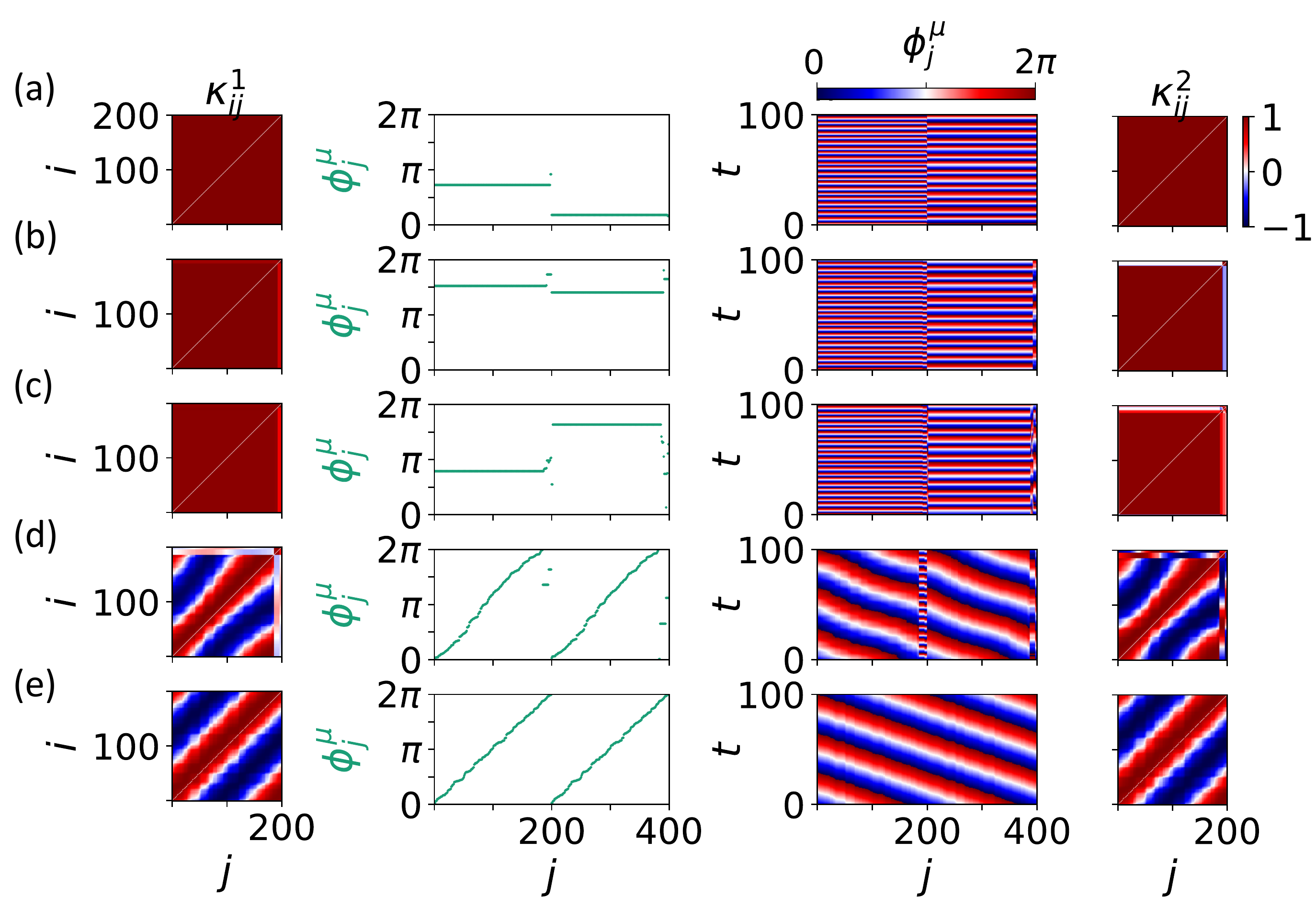}
\caption{Details of dynamical scenarios for healthy states without clusters (a),(b),(e) and pathological states with multifrequency clusters (c),(d) for different values of $\beta$ and $r$ chosen as in Fig.\,\ref{Fig2}. The left and right columns show snapshots of cytokine matrices $\kappa^1_{ij}$ (parenchymal layer) and $\kappa^2_{ij}$ (immune layer), respectively (color coded). Second column: snapshots of phases $\phi_j^\mu$ with the parenchymal nodes labeled $j=1,...,200$, and the immune nodes labeled $j=201,...,400$. Third column: space-time plot of phases $\phi_j^{\mu}(t)$ (color coded). Parameters: $\beta=0.45\pi$ and $r=1\%$ (a), $\beta=0.55\pi$ and $r=4\%$ (b), $\beta=0.60\pi$ and $r=4\%$ (c), $\beta=0.66\pi$ and $r=7.5\%$ (d), and $\beta=0.66\pi$ and $r=0\%$ (e). Within each layer $\mu$ the nodes are sorted first by $\langle \dot{\phi}_j^1 \rangle$, then by $\phi_j^1$, respectively. Other parameters as in Fig.\,\ref{Fig2}.
}
\label{Fig3}
\end{figure}

Figure~\ref{Fig3} shows details of these scenario for the same values of age parameter $\beta$ and tumor size $r$. The left and right columns show snapshots of the cytokine matrices $\kappa^1_{ij}$ (parenchymal layer) and $\kappa^2_{ij}$ (immune layer), respectively. The second column shows snapshots of the instantaneous phases $\phi_j^\mu$, and the third column depicts space-time plots of the phases $\phi_j^{\mu}(t)$ visualizing the oscillations. In the healthy state in Fig.\,\ref{Fig3}(a), (b) the cytokine matrices are uniform and temporally constant within each cluster. The snapshots of the phases $\phi_j^\mu$ (second column) and the space-time plots $\phi_j^\mu(t)$ (third column) show homogeneous in-phase oscillations in the parenchyma and the immune layer, respectively, though with different collective frequencies. In the pathological state in Fig.\,\ref{Fig3}(c) the small pathological cluster breaks off at $j=193, \ldots, 200$. For even larger age parameter $\beta=0.66\pi$ and tumor size $r=7.5\%$ in Fig.\,\ref{Fig3}(d) the phase dynamics changes qualitatively. We obtain a pathological two-frequency cluster with strong frequency difference, where the healthy part is no longer in-phase synchronized, but becomes a splay state. With increasing age parameter $\beta$ the frequency difference of the healthy cluster and the tumor cluster becomes larger, i.e., the tumor cells become autonomous. Thus larger age parameter can be associated with less favorable conditions of tumor disease (higher age, higher tumor malignity). A different scenario in dependence on the tumor size $r$ with increased age parameter $\beta$ leads from a mixed 2-frequency cluster to a healthy state, where the healthy cluster is a splay state (Fig.\,\ref{Fig3}(e)). This state is different from the frequency clusters in Fig.\,\ref{Fig3}(a)-(c), where each frequency cluster is phase synchronized. In Fig.\,\ref{Fig3}(d),(e) the frequency of the healthy cluster in the parenchymal layer is fixed equal to that of the immune layer, but the phases of the parenchyma can no longer be kept in-phase, thus a strong overall perturbation of the healthy cluster is visible. The frequency of the pathological cluster is much more strongly separated from that of the healthy cluster. When the tumor size $r$ is reduced for low age parameter $\beta$, below a certain $r$ the splay state does no longer occur and the in-phase state is recovered Fig.\,\ref{Fig3}(a). In contrast, for high age parameter $\beta$ the splay state occurs even without tumor and multifrequency clustering in a healthy state (see Fig.\,\ref{Fig3}(e)). The two different pathological scenarios in Fig.\,\ref{Fig3}(c) and (d) are connected with different responses of the immune layer and might indicate different malignity of the tumor.

\begin{figure}[ht]
\centering
\includegraphics[width = 1.0\textwidth]{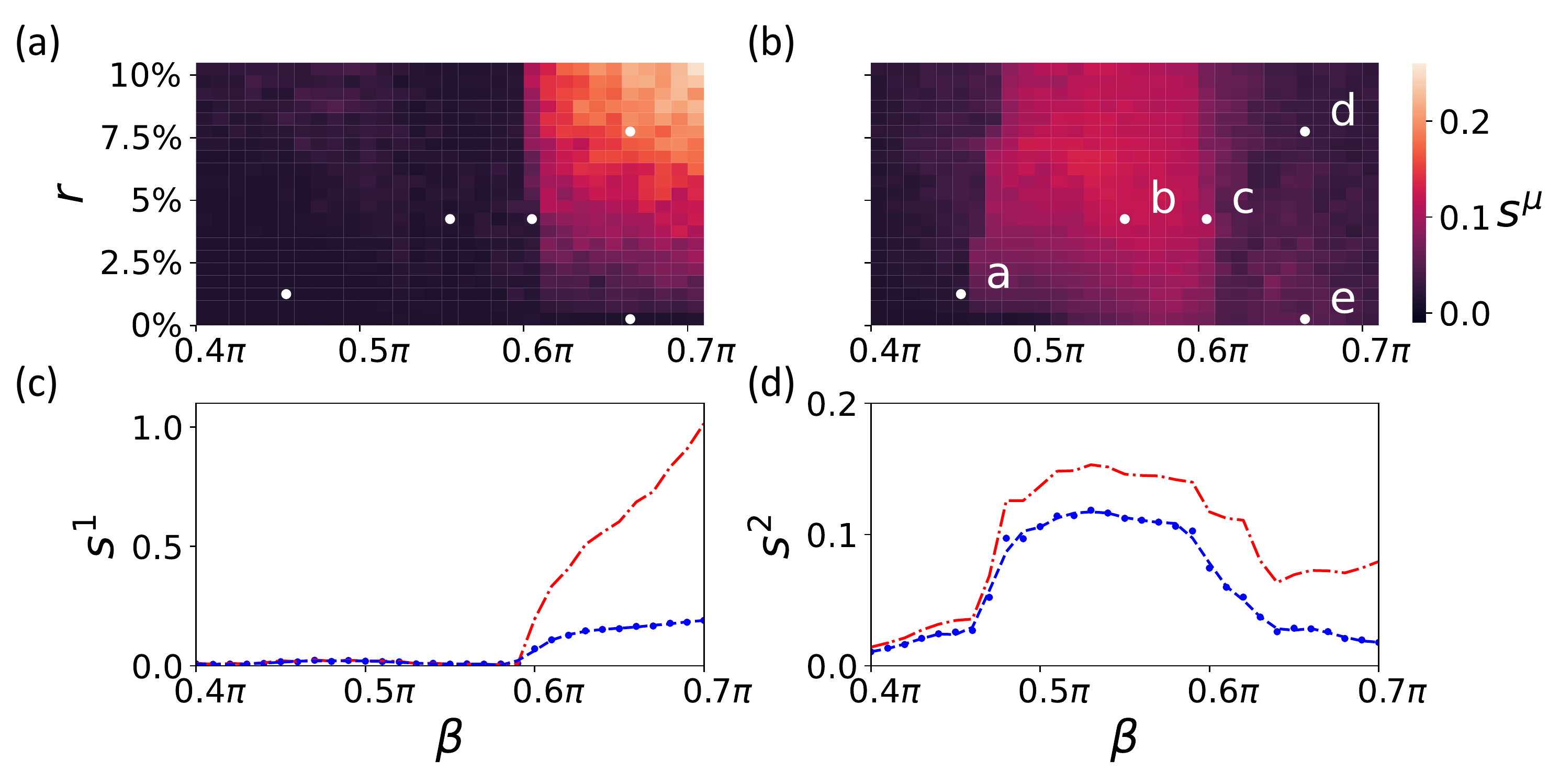}
\caption{Map of regimes: ensemble average $s^\mu$ of the standard deviation of the mean phase velocities in the parameter plane of age parameter $\beta$ and tumor size $r$) (ensemble size is $N_E=50$) for the parenchymal (a) and immune layer (b), respectively. Bright colors correspond to the formation of multifrequency clusters. The dynamics of the points marked (a)-(e) in the upper right panel are shown in Figs.\,\ref{Fig2} and \ref{Fig3}. The lower panels depict the ensemble average $s^\mu$ (blue) in dependence of $\beta$ for a fixed value of $r=7.5\%$ for the parenchymal (c) and immune layer (d), respectively. The dashed line is a regression curve for an ensemble size of $N_E=200$. For comparison the ensemble average $s^\mu$ of the standard deviation (blue curves) and the normalized standard deviation (red curves) of the spatially averaged mean phase velocities is shown in dependence of $\beta$ for the parenchymal (c) and immune layer (d), respectively. Other simulation parameters as in Fig.\,\ref{Fig2}.}
\label{Fig4}
\end{figure}

\begin{figure}[ht]
	\centering
	\includegraphics[width = .65\textwidth]{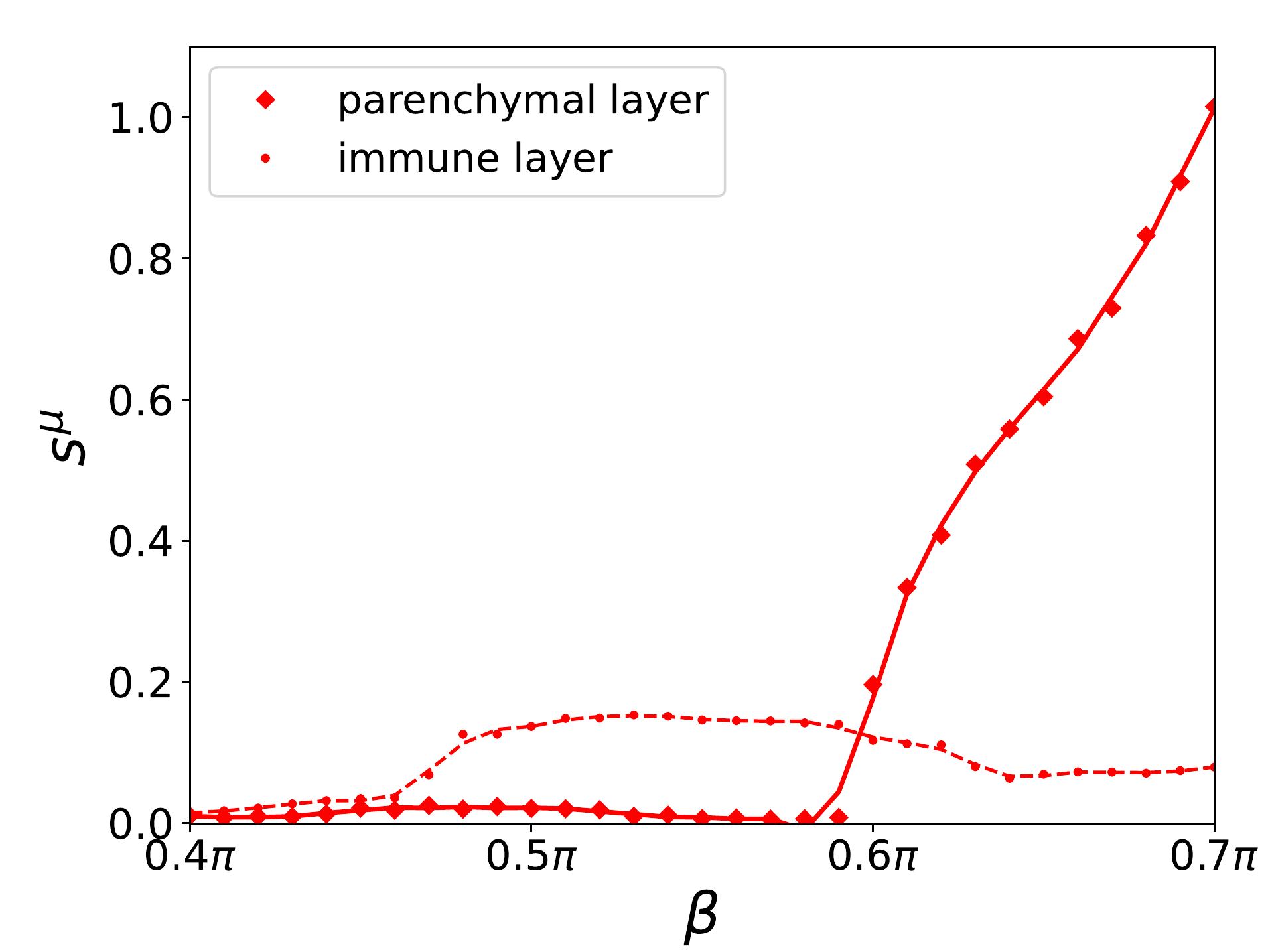}
	\caption{Ensemble average $s^\mu$ of the normalized standard deviation of the mean phase velocities in dependence of $\beta$ for a fixed value of $r=7.5\%$ for the parenchymal (solid curve, diamonds) and immune layer (broken curve, dots), respectively for an ensemble size of $N_E=200$. A cubic Savitzky–Golay filter has been applied to smooth the curves. Other simulation parameters as in Fig.\,\ref{Fig2}.}
	\label{Fig5}
\end{figure}

A map of regimes in the parameter plane of ($\beta,r$) is shown in Fig.\,\ref{Fig4}(a),(b), where the ensemble average $s^\mu$ of the standard deviation of the mean phase velocities is plotted for the parenchymal and immune layer, respectively. Bright colors correspond to large deviations of the frequencies (mean phase velocities), and hence to pathological two-cluster states. Their enhancement with increasing age parameter and tumor size can be clearly seen. In Figure~\ref{Fig4}(c),(d), a comparison between the ensemble average of the standard deviation and the normalized standard deviation is depicted. In comparison to the standard deviation, the normalized standard deviation shows higher values of the ensemble average $s^1$ in case of splay states (see Fig.\,\ref{Fig4}(c) for $\beta>0.6 \pi$). On the contrary, the small bulge of $s^1$ at $\beta=0.5 \pi$ is less pronounced. Figure~\ref{Fig5} depicts the comparison between the normalized standard deviation of the parenchymal and immune layer. It clearly shows the strong increase of the frequency deviation in the parenchyma associated with the pathogenicity of the tumor with increasing age parameter, while the much smaller bulge in the immune layer is associated with the immune layer activation before the tumor becomes clearly visible.

%-------------------------
% Sepsis
%-------------------------
\section{Sepsis}\label{sec:sepsis}

In our model we focus on the stage of sepsis which is characterized by generalized inflammation. Depending on the individual subject and the local cellular situation of cytokine and fluid influx, either de-escalation of the inflammatory reaction and restoration of homeostasis, or organ failure occurs. Sepsis is usually preceded by a pre-septic perturbation of the parenchyma, e.g., by a wound which is infected by germs. This perturbation is terminated after a while by blocking off the wound by blood coagulation and eventually healing. Under ``normal" conditions, the system returns to a healthy state; however, if sepsis occurs, an inflammatory immune response triggered by the infection spreads across the whole body (cytokine storm). As a consequence, the immune activity may invade large parts of the body through blood vessels and lead to severe organ failure and death. Whether sepsis terminates in a septic shock with severe consequences for the patient depends crucially on the ability of the immune and parenchymal system to regain homeostasis. As in the case of the tumor disease, the ability for returning to a healthy state, i.e., a frequency-synchronized state, is subsequently analyzed using the systemic sum parameter (age parameter) $\beta$.

For the simulations, we assume that a dysregulation of the cytokine activity in the immune system has already occurred due to a systemic immune response. The dysregulation of the cytokine activity is modeled by an initial perturbation of the cytokine activity matrix $\kappa^2_{ij}$ of the immune layer. For the latter, we consider a separation into two clusters with high activity between nodes from the same cluster ($\kappa^2_{ij}=1$) and no activity between nodes from different clusters ($\kappa^2_{ij}=0$). All other initial conditions are chosen randomly, as in the simulations for tumor disease in Sec.\,\ref{sec:tumor}. An example initial condition is presented in Fig.\,\ref{Fig9} in the Appendix. Note that we arbitrarily fix the cluster sizes of the initial cytokine activities in the immune layer for the rest of this paper. Furthermore, we increase the value of the interlayer coupling strength $\sigma$ compared to our simulations of the tumor disease. This choice is natural in order to understand the mechanism in action during the progression of sepsis. As described in Sec.\,\ref{sec:physiology}, pro-inflammatory cytokines act on endothelial cells and hence cause an increased blood vessel leakiness. As a result, more immune cells and cytokines enter the stroma which in consequence increases the immune-parenchymal interaction. 

In the following, we present simulation results for different choices of the age parameter $\beta$ showing that after an initial cytokine perturbation in the immune layer either the healthy frequency-synchronized state is restored, or the whole system goes to a desynchronized or multifrequency cluster state.

\begin{figure}[ht]
	\centering
	\includegraphics[width = .85\textwidth]{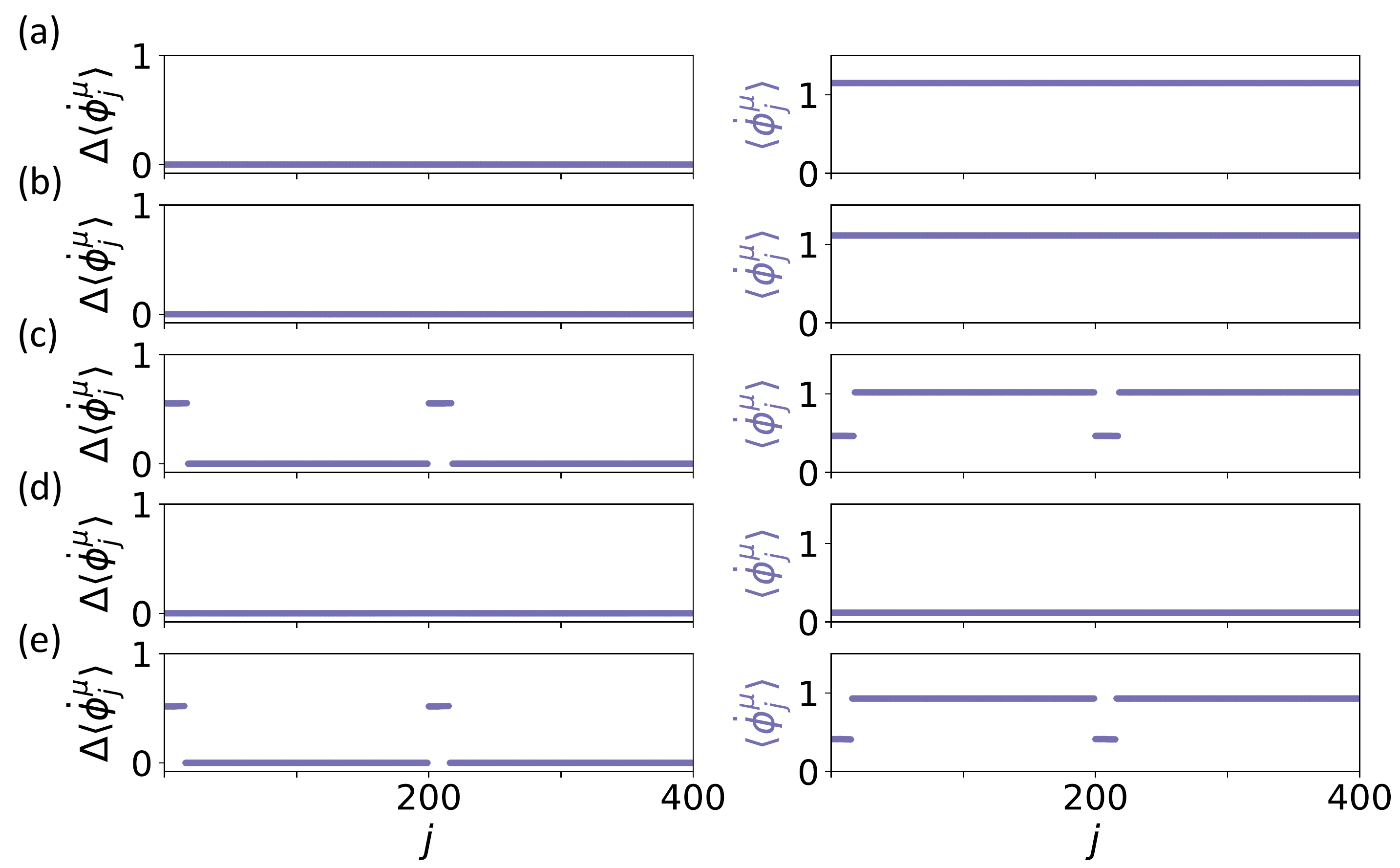}
	\caption{Dynamical scenarios for healthy states without clusters and pathological states with multifrequency clusters for different values of $\beta$. The left panels show the absolute differences of the mean phase velocities $\Delta\langle\dot{\phi}_j^\mu\rangle =|\langle \dot{\phi}_j^{1/2} \rangle-\langle \dot{\phi}_{100/300}^{1/2} \rangle|$. The right panels show mean phase velocities $\langle \dot{\phi}_j^\mu \rangle$. The parenchymal nodes are labeled by $j=1,...,200$, and the immune nodes by $j=201,...,400$. The age parameter is $\beta=0.5\pi$ (a), $\beta=0.6\pi$ (b), $\beta=0.6\pi$ (c), $\beta=0.7\pi$ (d) and $\beta=0.7\pi$ (e), where panels (b) and (c), and panels (d) and (e), respectively, only differ in the random initial conditions $\kappa^1_{ij}(0)$, $\phi_j^\mu(0)$. Within each layer $\mu$ the nodes are sorted first by $\langle \dot{\phi}_j^1 \rangle$, then by $\phi_j^1$. Simulation parameters: $\sigma=1$, $\omega^1=\omega^2=0$. Other parameters as in Fig.\,\ref{Fig2} and the initial conditions as in Fig.\,\ref{Fig9}.}
	\label{Fig6}
\end{figure}
\begin{figure}[ht]
	\centering
	\includegraphics[width = .9\textwidth]{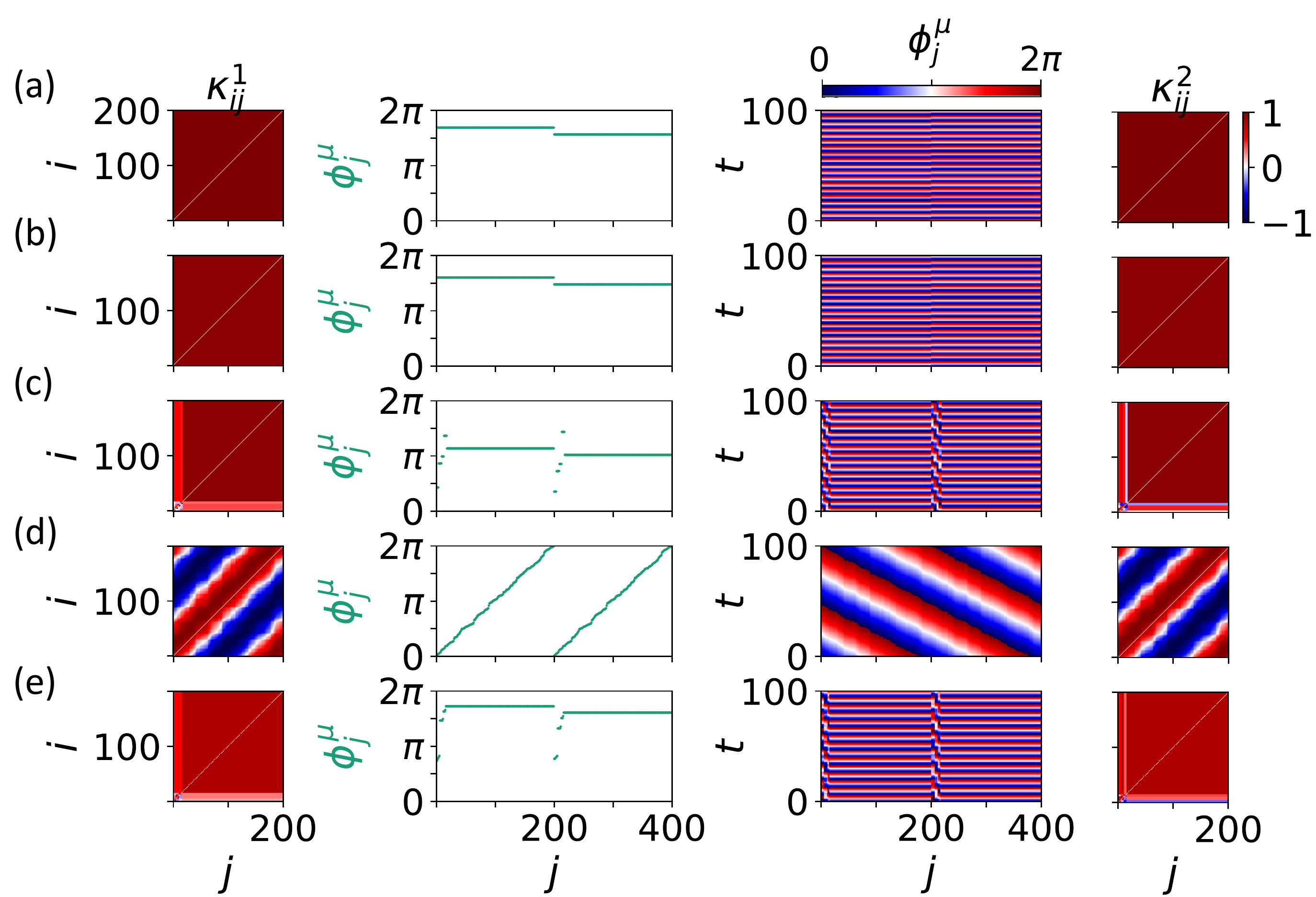}
	\caption{Details of dynamical scenarios for healthy states without clusters (a),(b),(d) and pathological states with multifrequency clusters (c),(e) for different values of $\beta$ chosen as in Fig.\,\ref{Fig6}. The left and right columns show snapshots of cytokine activity matrices $\kappa^1_{ij}$ (parenchymal layer) and $\kappa^2_{ij}$ (immune layer), respectively (color coded). Second column: snapshots of phases $\phi_j^\mu$ with the parenchymal nodes labeled $j=1,...,200$, and the immune nodes labeled $j=201,...,400$. Third column: space-time plot of phases $\phi_j^{\mu}(t)$ (color coded). The parameters $\beta$ and all other parameters are chosen as in Fig.\,\ref{Fig6}.
	}
	\label{Fig7}
\end{figure}

Figure \ref{Fig6} shows five representative dynamical scenarios that are induced by pathological initial cytokine activity in the immune layer. For an age parameter $\beta=0.5\pi$ in Fig.\ref{Fig6}(a), we observe that the system relaxes to a healthy state, i.e., frequency synchronized state, after an initial immune layer perturbation. Hence, the pro-inflammatory response is stopped. A transient of $1000$ time units is discarded before the temporal average over $T=1000$ is taken in the mean phase velocities $\langle \dot{\phi}_j^\mu \rangle$.  In addition, Fig.\,\ref{Fig7}(a) shows that in the final state the phases in both layers are in-phase synchronized representing a resilient healthy state. The same scenario can be observed as well for an increased value of the age parameter $\beta=0.6\pi$ in Figs.\,\ref{Fig6}(b) and \ref{Fig7}(b). While for $\beta=0.5\pi$ it is unlikely to obtain a pathological state, i.e., a desynchronized state, from any random initial conditions of $\kappa^1_{ij}$ and the phases $\phi_j^\mu$, the probability of a pathological state for an ensemble of random initial conditions $\kappa^1_{ij}(0)$, $\phi_j^\mu(0)$ increases for $\beta=0.6\pi$. In Figures~\ref{Fig6}(c) and \ref{Fig7}(c), we depict a dynamical scenario where the initial immune layer perturbation induces a desynchronization of the parenchymal layer. We clearly observe the presence of a two-frequency cluster. In this situation, the initial activated immune response can not be compensated by the coupled system and pushes the parenchyma away from a homeostatic state that may have severe consequences for the organic tissue, compare with the discussion for tumor disease in Sec.\,\ref{sec:tumor}. For even higher values of the age parameter $\beta=0.7\pi$ the probability of desynchronization increases. Depending upon the random initial conditions of $\kappa^1_{ij}$ and the phases $\phi_j^\mu$, also for this value of $\beta$, we may obtain a frequency-synchronized (healthy) or a desynchronized (pathological) state. In Figures~\ref{Fig6}(d) and \ref{Fig7}(d), we display a frequency-synchronized state for $\beta=0.7\pi$. However, this particular state is not in-phase synchronized any more, as it is the case for the healthy states in Figure~\ref{Fig7}(a),(b). Figure~\ref{Fig7}(d) shows this state which possesses a splay distribution of the phases. Splay distributions of the phases may be interpreted as more vulnerable and less resilient against further perturbations by pathological cells in the parenchymal layer or a "second hit" phenomena known for sepsis. The more likely dynamical scenario of an emergent pathological two-frequency cluster for the same value of $\beta=0.7\pi$ is displayed in Figs.\,\ref{Fig6}(e) and \ref{Fig7}(e).

For all presented scenarios in Figs.\,\ref{Fig6} and \ref{Fig7}, we note that the frequencies of the parenchymal and the immune layer are locked. The emergence of the dynamical phenomenon of locking~\cite{PIK01} can be explained by the increased value of the interlayer coupling strength $\sigma$ compared to the simulation of the tumor disease, see Sec.\,\ref{sec:tumor}. Locking can be interpreted as the dynamical manifestation of the physiological observation that the immune system is taking control of the whole system in case of sepsis. For smaller interlayer coupling strength this locking cannot be observed, and hence desynchronization induced by a pathologically activated immune system is not possible.

\begin{figure}[ht]
	\centering
	\includegraphics[width = .65\textwidth]{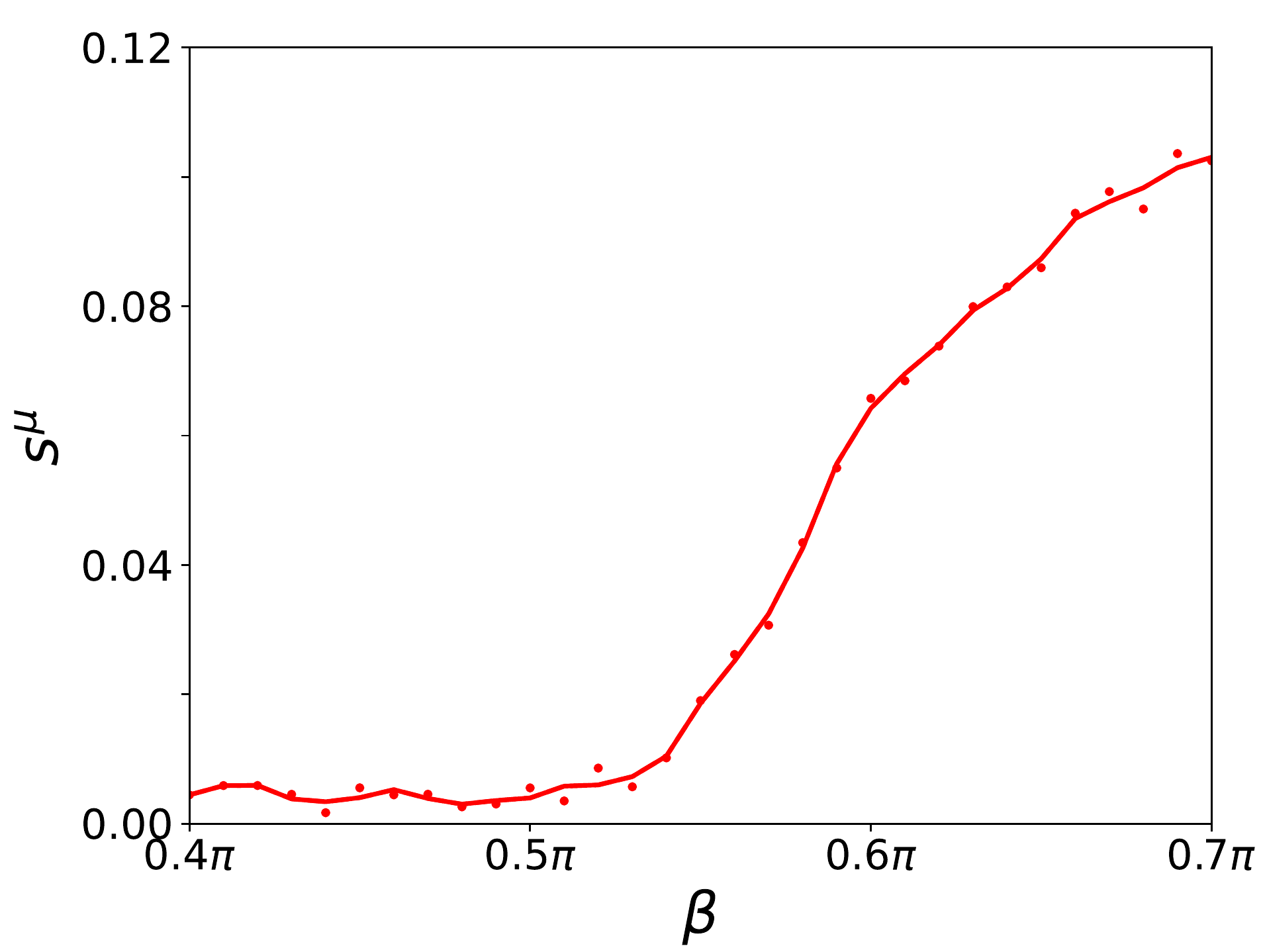}
	\caption{Ensemble average $s^\mu$ of the normalized standard deviation of the mean phase velocities in dependence of $\beta$ for a fixed value of $\sigma=1$ for the parenchymal and immune layer (both curves coincide) for an ensemble size of $N_E=200$. A cubic Savitzky-Golay filter has been applied to smooth the curve. Other simulation parameters as in Fig.\,\ref{Fig6}.}
	\label{Fig8}
\end{figure}

The dynamical scenarios discussed in this section are very similar to the observations presented for tumor disease. Additionally, in agreement with the results found for tumor disease, the desynchronization of the parenchymal layer due to an initially activated immune layer is observed for higher age parameter $\beta$ with increasing probability. We note that also here resilient and vulnerable healthy states may coexist with pathological desynchronized states. The increased desynchronization probability with increasing $\beta$ is reflected in the increasing ensemble average of the normalized standard deviation of the mean phase velocities in Fig.\,\ref{Fig8}. 

\section{Conclusion}\label{sec:conclusion}
In this article, we have proposed a two-layer network model for carcinogenesis and sepsis based upon the interaction of parenchymal cells and immune cells via cytokines and the co-evolutionary dynamics of parenchymal, immune cells and cytokines. Certain parallels between cancer and infectious disease have been unveiled in the medical sciences, however, little is understood about the underlying mechanism behind these very similar pathological states induced by tumor and sepsis, respectively~\cite{HOT14}. With this study, we propose a novel paradigm of unified functional modeling of tumor disease and sepsis from the complex dynamical network perspective by choosing the nonspecific innate immune system as the reference point for both diseases. The proposed model is not a detailed model of organs but a functional model of dynamic interactions. Here, cytokine activity is described as information flow within and between the parenchymal and the immune layer. In particular, the communication between cells of the parenchyma, and between cells of the immune layer is modeled by adaptive coupling weights. This approach is complementary to works modeling the cytokine concentrations as additional dynamical species~\cite{YIU12}. Thus our perspective accounts for the cytokine activation rather than their physical mass.

In this paper we have presented the simplest form of a two-layer network model based upon adaptively coupled phase oscillators. Although many simplifying assumptions have been made, this model can already capture essential unifying features of tumor disease and sepsis. Two important tunable model parameters have been identified, i.e., the tumor size $r$ denoting the fraction of mutant (pathological) parenchymal cells which have a deviating natural frequency of single-cell basic metabolic activity, and an adaptivity parameter $\beta$ (called age parameter) summarizing various physiological conditions like age, inflammaging, adiposity, pre-existing illness, physical inactivity, nutritional influence, and malignancy in case of tumor disease. The healthy system is modeled as a completely frequency-synchronized homeostatic state of metabolic activity in both the parenchyma and the immune layer, while in the pathological case the parenchyma splits into two frequency clusters with different collective frequency, one corresponding to the healthy part, and one corresponding to organ failure. The desynchronization of the parenchymal layer is observed for higher age parameter $\beta$ with increasing probability both in case of tumor disease and sepsis. 

To characterize the dynamical patterns resulting from this model, we have introduced the temporally averaged mean phase velocities of the cells as collective frequencies of metabolism, and their ensemble-averaged standard deviation $s^\mu$ as a measure of pathogenicity in case of the parenchymal layer ($\mu=1$), and as a measure of metabolic activation in case of the immune layer ($\mu=2$). Thus $s^1\neq 0$ is an immediate indicator of a pathological state. Further, we have discussed the initial perturbation for both types of pathological conditions, i.e., tumor and sepsis. For the analysis of the tumor disease, we assume heterogeneity in the natural frequency distribution of the parenchymal layer to account for the fact that mutated parenchymal cells may possess a faster metabolism than healthy parenchymal cells. For sepsis, we consider, however, a homogeneous network of oscillator with an initially activated cytokine matrix representing a systemic immune response. Moreover, for sepsis, we assume a high coupling strength between the parenchymal and the immune layer caused by the pro-inflammatory immune response.

In case of tumor disease, our simulations show that for small values of the age parameter $\beta$, as representative for young and healthy subjects, almost all simulations lead to a healthy state even for a relatively high fraction $r$ of mutated cells. Thus the system is robust against random mutations of parenchymal cells. For increasing age parameter, the immune layer turns out to be more activated in order to keep the parenchymal layer synchronized. Here, only for a low fraction of mutated cells $r$ the healthy state can be maintained. For higher values of the fraction $r$ the immune layer is not able anymore to keep the parenchyma synchronized, and a pathological desynchronized state emerges. For high values of the age parameter the immune layer is completely dysregulated, and a severe pathological state of tumor disease is almost inevitable. One might speculate that in the in-phase scenario the pathological cluster (tumor) may be removed (by surgery) so that the healthy state is restored, whereas in the splay scenario the non-tumor cluster is already so strongly perturbed that removal of the tumor may not bring the system back to the healthy state.

In analogy with the study of tumor disease, we have demonstrated various dynamical scenarios for different age parameters in case of sepsis. Also here, the age parameter $\beta$ has been shown to be critical for the system to return to the healthy state. For specifically chosen parameter values, close to those chosen for the examples of tumor disease, we have visualized the emergence of different synchronized and desynchronized states. Furthermore, we have found parameter regimes for which healthy states may coexist with pathological desynchronized states, i.e., it depends upon the random initial conditions which of the two states is asymptotically reached. Further systematic studies should investigate the detailed dependence upon the initial perturbation of the cytokin activity matrix in the immune layer, i.e., the initial cluster and its size, and upon the choice of the interlayer coupling strength. We hypothesize that the interlayer coupling strength $\sigma$ may be used to model functionally the progression of sepsis where an increased $\sigma$ may be interpreted as an ongoing in-stream of cytokines from the blood vessels into the stroma. For both tumor disease and sepsis our model represents only a first step towards more detailed and systematic investigations.

%in summary, the functional proposed model is capable of capturing the main healthy and pathological conditions observed in clinical studies for tumors and sepsis. Moreover, the dynamical scenarios discussed in this article for sepsis and tumor disease are closely related. Remarkably, for both pathological conditions considered in this work, the desynchronization of the parenchymal layer is observed only for lower age parameters with increasing likeliness. Due to these findings, we believe that the two-layer adaptive dynamical model may be an important first step to shed light on the basic mechanisms responsible for the evident parallels between tumor disease and sepsis.\\

In summary, a unified disease model has been established using the innate immune system as the reference point. Diseases such as tumorigenesis and sepsis have been modeled by studying the dynamic functional interactions between parenchyma and stroma. For this purpose, a unified system of dynamic equations has been developed, in which only three different parameters, the age parameter, the number of mutated cells, and the initial cytokine activation are varied. Thereby, carcinogenesis, organ dysfunction in sepsis, and recurrence risk can be described in a correct functional context. Our studies open a perspective for further dynamic disease modeling. For instance, more detailed models, including disease progression may be developed in future studies.

\section*{Appendix: Initial conditions for sepsis}
\begin{figure}[ht]
	\centering
	\includegraphics[width = .75\textwidth]{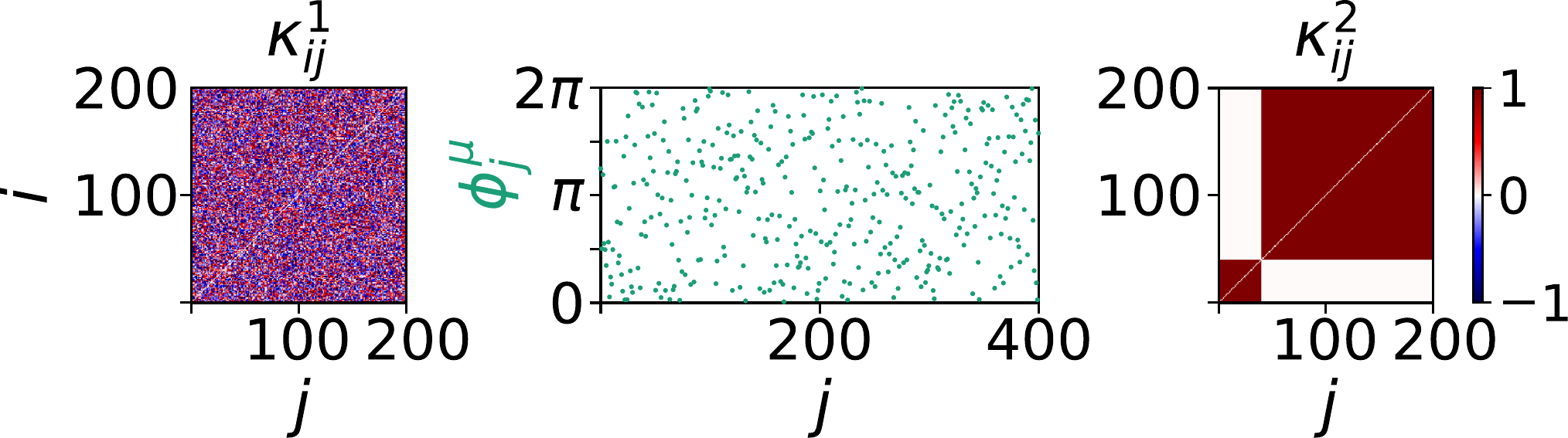}
	\caption{Initial conditions of sepsis: Cytokine dysregulation expressed by a cluster structure of the cytokine activity matrix $\kappa^2_{ij}$ imposes a systemic activation of the immune layer representing the beginning of sepsis. The figure shows a sample initial condition used for simulations of Eqs.~\eqref{eq:DGL_somatic}--\eqref{eq:DGL_immune} with $N=200$. The initial cytokine activities in the parenchymal layer $\kappa_{ij}^1$ and the initial phases in both layers are randomly drawn from a uniform distribution on the interval $[-1,1]$ and the interval $[0, 2\pi]$, respectively. The cytokine activities in the immune layer $\kappa_{ij}^2$ are initially given by a two-cluster structure where the smallest cluster has size $40$. The cytokine activities $\kappa_{ij}^2$ are $1$ within and $0$ between the clusters.
	}
	\label{Fig9}
\end{figure}

\section*{Conflict of Interest Statement}
The authors declare that the research was conducted in the absence of any commercial or financial relationships that could be construed as a potential conflict of interest.

\section*{Author Contributions}
JS and RB performed the numerical simulations and analyzed the simulated data. The idea of unified treatment of tumor disease and sepsis was due to TL. The idea of an adaptive two-layer network model where the cytokine activity is associated with adaptive coupling weights was due to ES. All authors designed the study and contributed to the preparation of the manuscript. All the authors have read and approved the final manuscript.

\section*{Funding}
This work was supported by the Deutsche Forschungsgemeinschaft (DFG, German Research Foundation, project Nos. 429685422 and 440145547) and the Open Access Publication Fund of TU Berlin.
%\section*{Acknowledgments}

%\bibliographystyle{frontiersinSCNS_ENG_HUMS} % for Science, Engineering and Humanities and Social Sciences articles, for Humanities and Social Sciences articles please include page numbers in the in-text citations
%\bibliographystyle{frontiersinHLTH&FPHY} % for Health, Physics and Mathematics articles

%\bibliography{ref}

\begin{thebibliography}{120}
\providecommand{\natexlab}[1]{#1}
\expandafter\ifx\csname urlstyle\endcsname\relax
  \providecommand{\doi}[1]{doi:\discretionary{}{}{}#1}\else
  \providecommand{\doi}{doi:\discretionary{}{}{}\begingroup
  \urlstyle{rm}\Url}\fi
\providecommand{\selectlanguage}[1]{\relax}
\providecommand{\bibAnnoteFile}[1]{%
  \IfFileExists{#1}{\begin{quotation}\noindent\textsc{Key:} #1\\
  \textsc{Annotation:}\ \input{#1}\end{quotation}}{}}
\providecommand{\bibAnnote}[2]{%
  \begin{quotation}\noindent\textsc{Key:} #1\\
  \textsc{Annotation:}\ #2\end{quotation}}

\bibitem[{Abbott and Nelson(2000)}]{ABB00}
Abbott, L.~F. and Nelson, S. (2000).
\newblock Synaptic plasticity: taming the beast.
\newblock \emph{Nat. Neurosci.} 3, 1178--1183.
\newblock \doi{10.1038/81453}
\input{ABB00}

\bibitem[{Abrams and Strogatz(2004)}]{ABR04}
Abrams, D.~M. and Strogatz, S.~H. (2004).
\newblock Chimera states for coupled oscillators.
\newblock \emph{Phys. Rev. Lett.} 93, 174102.
\newblock \doi{10.1103/physrevlett.93.174102}
\input{ABR04}

\bibitem[{Acebr{\'o}n et~al.(2005)Acebr{\'o}n, Bonilla, P\'{e}rez~Vicente,
  Ritort, and Spigler}]{ACE05}
Acebr{\'o}n, J.~A., Bonilla, L.~L., P\'{e}rez~Vicente, C.~J., Ritort, F., and
  Spigler, R. (2005).
\newblock The {K}uramoto model: A simple paradigm for synchronization
  phenomena.
\newblock \emph{Rev. Mod. Phys.} 77, 137--185.
\newblock \doi{10.1103/revmodphys.77.137}
\input{ACE05}

\bibitem[{Albert and Barab{\'a}si(2002)}]{ALB02a}
Albert, R. and Barab{\'a}si, A.~L. (2002).
\newblock Statistical mechanics of complex networks.
\newblock \emph{Rev. Mod. Phys.} 74, 47--97.
\newblock \doi{10.1103/revmodphys.74.47}
\input{ALB02a}

\bibitem[{Altan-Bonnet and Mukherjee(2019)}]{ALT19}
Altan-Bonnet, G. and Mukherjee, R. (2019).
\newblock Cytokine-mediated communication: a quantitative appraisal of immune
  complexity.
\newblock \emph{Nat. Rev. Immunol.} 19, 205--217.
\newblock \doi{10.1038/s41577-019-0131-x}
\input{ALT19}

\bibitem[{Andrzejak et~al.(2017)Andrzejak, Ruzzene, and Malvestio}]{AND17}
Andrzejak, R.~G., Ruzzene, G., and Malvestio, I. (2017).
\newblock Generalized synchronization between chimera states.
\newblock \emph{Chaos} 27, 053114.
\newblock \doi{10.1063/1.4983841}
\input{AND17}

\bibitem[{Aoki(2015)}]{AOK15}
Aoki, T. (2015).
\newblock Self-organization of a recurrent network under ongoing synaptic
  plasticity.
\newblock \emph{Neural Netw.} 62, 11--19.
\newblock \doi{10.1016/j.neunet.2014.05.024}
\input{AOK15}

\bibitem[{Aoki and Aoyagi(2009)}]{AOK09}
Aoki, T. and Aoyagi, T. (2009).
\newblock Co-evolution of phases and connection strengths in a network of phase
  oscillators.
\newblock \emph{Phys. Rev. Lett.} 102, 034101.
\newblock \doi{10.1103/physrevlett.102.034101}
\input{AOK09}

\bibitem[{Arends et~al.(2015)Arends, Bertz, Bischoff, Fietkau, Herrmann, Holm
  et~al.}]{ARE15}
Arends, J., Bertz, H., Bischoff, S.~C., Fietkau, R., Herrmann, H.~J., Holm, E.,
  et~al. (2015).
\newblock Klinische ern{\"a}hrung in der onkologie.
\newblock \emph{Aktuel. Ernahrungsmed.} 40, 1.
\newblock \doi{10.1055/s-0035-1552741}
\input{ARE15}

\bibitem[{Bartsch et~al.(2015)Bartsch, Liu, Bashan, and Ivanov}]{BAR15b}
Bartsch, R.~P., Liu, K. K.~L., Bashan, A., and Ivanov, P.~C. (2015).
\newblock Network {P}hysiology: {H}ow {O}rgan {S}ystems {D}ynamically
  {I}nteract.
\newblock \emph{PLoS One} 10, 11.
\newblock \doi{10.1371/journal.pone.0142143}
\input{BAR15b}

\bibitem[{Bartsch et~al.(2012)Bartsch, Schumann, Kantelhardt, Penzel, and
  Ivanov}]{BAR12e}
Bartsch, R.~P., Schumann, A.~Y., Kantelhardt, J.~W., Penzel, T., and Ivanov,
  P.~C. (2012).
\newblock Phase transitions in physiologic coupling.
\newblock \emph{Proc. Natl. Acad. Sci. U.S.A.} 109, 10181.
\newblock \doi{https://doi.org/10.1073/pnas.1204568109}
\input{BAR12e}

\bibitem[{Bashan et~al.(2012)Bashan, Bartsch, Kantelhardt, Havlin, and
  Ivanov}]{BAS12b}
Bashan, A., Bartsch, R.~P., Kantelhardt, J.~W., Havlin, S., and Ivanov, P.~C.
  (2012).
\newblock Network physiology reveals relations between network topology and
  physiological function.
\newblock \emph{Nat. Commun.} 31, 702.
\newblock \doi{10.1038/ncomms1705}
\input{BAS12b}

\bibitem[{Beneke(1971)}]{BEN71}
Beneke, G. (1971).
\newblock \emph{Altersabh{\"a}ngige Ver{\"a}nderung des Kollagens und der
  Bindegewebszellen} (Schattauer Verlag), chap.~1.
\newblock Altern und Entwicklung. 1--37
\input{BEN71}

\bibitem[{Berner(2021)}]{BER21c}
Berner, R. (2021).
\newblock \emph{{Patterns of Synchrony in Complex Networks of Adaptively
  Coupled Oscillators}}.
\newblock Springer Theses (Cham: Springer).
\newblock \doi{10.1007/978-3-030-74938-5}
\input{BER21c}

\bibitem[{Berner et~al.(2019{\natexlab{a}})Berner, Fialkowski, Kasatkin,
  Nekorkin, Yanchuk, and Sch{\"o}ll}]{BER19a}
Berner, R., Fialkowski, J., Kasatkin, D.~V., Nekorkin, V.~I., Yanchuk, S., and
  Sch{\"o}ll, E. (2019{\natexlab{a}}).
\newblock Hierarchical frequency clusters in adaptive networks of phase
  oscillators.
\newblock \emph{Chaos} 29, 103134.
\newblock \doi{10.1063/1.5097835}
\input{BER19a}

\bibitem[{Berner et~al.(2021{\natexlab{a}})Berner, Mehrmann, Sch{\"o}ll, and
  Yanchuk}]{BER21}
Berner, R., Mehrmann, V., Sch{\"o}ll, E., and Yanchuk, S. (2021{\natexlab{a}}).
\newblock The multiplex decomposition: An analytic framework for multilayer
  dynamical networks.
\newblock \emph{SIAM J. Appl. Dyn. Syst.} 20, 1752--1772.
\newblock \doi{10.1137/21m1406180}
\input{BER21}

\bibitem[{Berner et~al.(2020{\natexlab{a}})Berner, Polanska, Sch{\"o}ll, and
  Yanchuk}]{BER20c}
Berner, R., Polanska, A., Sch{\"o}ll, E., and Yanchuk, S. (2020{\natexlab{a}}).
\newblock Solitary states in adaptive nonlocal oscillator networks.
\newblock \emph{Eur. Phys. J. Spec. Top.} 229, 2183--2203.
\newblock \doi{https://doi.org/10.1140/epjst/e2020-900253-0}
\input{BER20c}

\bibitem[{Berner et~al.(2020{\natexlab{b}})Berner, Sawicki, and
  Sch{\"o}ll}]{BER20}
Berner, R., Sawicki, J., and Sch{\"o}ll, E. (2020{\natexlab{b}}).
\newblock Birth and stabilization of phase clusters by multiplexing of adaptive
  networks.
\newblock \emph{Phys. Rev. Lett.} 124, 088301.
\newblock \doi{10.1103/physrevlett.124.088301}
\input{BER20}

\bibitem[{Berner et~al.(2019{\natexlab{b}})Berner, Sch{\"o}ll, and
  Yanchuk}]{BER19}
Berner, R., Sch{\"o}ll, E., and Yanchuk, S. (2019{\natexlab{b}}).
\newblock Multiclusters in networks of adaptively coupled phase oscillators.
\newblock \emph{SIAM J. Appl. Dyn. Syst.} 18, 2227--2266.
\newblock \doi{10.1137/18m1210150}
\input{BER19}

\bibitem[{Berner et~al.(2021{\natexlab{b}})Berner, Vock, Sch{\"o}ll, and
  Yanchuk}]{BER20b}
Berner, R., Vock, S., Sch{\"o}ll, E., and Yanchuk, S. (2021{\natexlab{b}}).
\newblock Desynchronization transitions in adaptive networks.
\newblock \emph{Phys. Rev. Lett.} 126, 028301.
\newblock \doi{10.1103/physrevlett.126.028301}
\input{BER20b}

\bibitem[{Berner et~al.(2021{\natexlab{c}})Berner, Yanchuk, Maistrenko, and
  Sch{\"o}ll}]{BER21e}
Berner, R., Yanchuk, S., Maistrenko, Y., and Sch{\"o}ll, E.
  (2021{\natexlab{c}}).
\newblock Generalized splay states in phase oscillator networks.
\newblock \emph{Chaos} 31, 073128.
\newblock \doi{10.1063/5.0056664}
\input{BER21e}

\bibitem[{Berner et~al.(2021{\natexlab{d}})Berner, Yanchuk, and
  Sch{\"o}ll}]{BER20a}
Berner, R., Yanchuk, S., and Sch{\"o}ll, E. (2021{\natexlab{d}}).
\newblock What adaptive neuronal networks teach us about power grids.
\newblock \emph{Phys. Rev. E} 103, 042315.
\newblock \doi{10.1103/physreve.103.042315}
\input{BER20a}

\bibitem[{Boccaletti et~al.(2014)Boccaletti, Bianconi, Criado, del Genio,
  G\'omez-Garde\~nes, Romance et~al.}]{BOC14}
Boccaletti, S., Bianconi, G., Criado, R., del Genio, C.~I., G\'omez-Garde\~nes,
  J., Romance, M., et~al. (2014).
\newblock The structure and dynamics of multilayer networks.
\newblock \emph{Phys. Rep.} 544, 1--122.
\newblock \doi{10.1016/j.physrep.2014.07.001}
\input{BOC14}

\bibitem[{Boccaletti et~al.(2018)Boccaletti, Pisarchik, del Genio, and
  Amann}]{BOC18}
Boccaletti, S., Pisarchik, A.~N., del Genio, C.~I., and Amann, A. (2018).
\newblock \emph{Synchronization: {F}rom Coupled Systems to Complex Networks}
  (Cambridge: Cambridge University Press)
\input{BOC18}

\bibitem[{Bomans et~al.(2018)Bomans, Schenz, Sztwiertnia, Schaack, Weigand, and
  Uhle}]{BOM18}
Bomans, K., Schenz, J., Sztwiertnia, I., Schaack, D., Weigand, M.~A., and Uhle,
  F. (2018).
\newblock Sepsis induces a long-lasting state of trained immunity in bone
  marrow monocytes.
\newblock \emph{Front. Immunol.} 9, 2685.
\newblock \doi{10.3389/fimmu.2018.02685}
\input{BOM18}

\bibitem[{Brunkhorst et~al.(2018)Brunkhorst, Weigand, Pletz, Gastmeier, Lemmen,
  Meier-Hellmann et~al.}]{BRU18b}
Brunkhorst, F.~M., Weigand, M., Pletz, M., Gastmeier, P., Lemmen,
  S.~W., Meier-Hellmann, A., et~al. (2018).
\newblock S3-Leitlinie Sepsis - Pr{\"a}vention, Diagnose, Therapie und
  Nachsorge
\input{BRU18b}

\bibitem[{Calder et~al.(2017)Calder, Bosco, Bourdet-Sicard, Capuron, Delzenne,
  Dor\'{e} et~al.}]{CAL17a}
Calder, P.~C., Bosco, N., Bourdet-Sicard, R., Capuron, L., Delzenne, N.,
  Dor\'{e}, J., et~al. (2017).
\newblock Health relevance of the modification of low grade inflammation in
  ageing (inflammageing) and the role of nutrition.
\newblock \emph{Ageing Res. Rev.} 40, 95.
\newblock \doi{10.1016/j.arr.2017.09.001}
\input{CAL17a}

\bibitem[{Caporale and Dan(2008)}]{CAP08a}
Caporale, N. and Dan, Y. (2008).
\newblock Spike timing-dependent plasticity: A {H}ebbian learning rule.
\newblock \emph{Annu. Rev. Neurosci.} 31, 25--46.
\newblock \doi{10.1146/annurev.neuro.31.060407.125639}
\input{CAP08a}

\bibitem[{Chen et~al.(2012)Chen, Liu, Liu, Li, and Aihara}]{CHE12a}
Chen, L., Liu, R., Liu, Z.~P., Li, M., and Aihara, K. (2012).
\newblock Detecting early-warning signals for sudden deterioration of complex
  diseases by dynamical network biomarkers.
\newblock \emph{Sci. Rep.} 2, 342
\input{CHE12a}

\bibitem[{Chen et~al.(2006)Chen, Hu, Stanley, Novak, and Ivanov}]{CHE06b}
Chen, Z., Hu, K., Stanley, H.~E., Novak, V., and Ivanov, P.~C. (2006).
\newblock Cross-correlation of instantaneous phase increments in pressure-flow
  fluctuations: Applications to cerebral autoregulation.
\newblock \emph{Phys. Rev. E} 73, 031915.
\newblock \doi{10.1103/physreve.73.031915}
\input{CHE06b}

\bibitem[{Chockley and Keshamouni(2016)}]{CHO16c}
Chockley, P.~J. and Keshamouni, V.~G. (2016).
\newblock Immunological consequences of epithelial-mesenchymal transition in
  tumor progression.
\newblock \emph{J. Immunol.} 197, 691.
\newblock \doi{https://doi.org/10.4049/jimmunol.1600458}
\input{CHO16c}

\bibitem[{Chovatiya and Medzhitov(2014)}]{CHO14d}
Chovatiya, R. and Medzhitov, R. (2014).
\newblock Stress, inflammation, and defense of homeostasis.
\newblock \emph{Mol. Cell} 54, 281.
\newblock \doi{10.1016/j.molcel.2014.03.030}
\input{CHO14d}

\bibitem[{Coussens and Werb(2002)}]{COU02}
Coussens, L.~M. and Werb, Z. (2002).
\newblock Inflammation and cancer.
\newblock \emph{Nature} 420, 860.
\newblock \doi{https://doi.org/10.1038/nature01322}
\input{COU02}

\bibitem[{Dahms et~al.(2012)Dahms, Lehnert, and Sch{\"o}ll}]{DAH12}
Dahms, T., Lehnert, J., and Sch{\"o}ll, E. (2012).
\newblock Cluster and group synchronization in delay-coupled networks.
\newblock \emph{Phys. Rev. E} 86, 016202.
\newblock \doi{10.1103/physreve.86.016202}
\input{DAH12}

\bibitem[{De~Domenico et~al.(2015)De~Domenico, Nicosia, Arenas, and
  Latora}]{DE15}
De~Domenico, M., Nicosia, V., Arenas, A., and Latora, V. (2015).
\newblock Structural reducibility of multilayer networks.
\newblock \emph{Nat. Commun.} 6, 6864.
\newblock \doi{10.1038/ncomms7864}
\input{DE15}

\bibitem[{De~Domenico et~al.(2013)De~Domenico, Sol{\'e}-Ribalta, Cozzo,
  Kivel{\"a}, Moreno, Porter et~al.}]{DE13}
De~Domenico, M., Sol{\'e}-Ribalta, A., Cozzo, E., Kivel{\"a}, M., Moreno, Y.,
  Porter, M.~A., et~al. (2013).
\newblock Mathematical formulation of multilayer networks.
\newblock \emph{Phys. Rev. X} 3, 041022.
\newblock \doi{10.1103/physrevx.3.041022}
\input{DE13}

\bibitem[{Drauschke et~al.(2020)Drauschke, Sawicki, Berner, Omelchenko, and
  Sch{\"o}ll}]{DRA20}
Drauschke, F., Sawicki, J., Berner, R., Omelchenko, I., and Sch{\"o}ll, E.
  (2020).
\newblock Effect of topology upon relay synchronization in triplex neuronal
  networks.
\newblock \emph{Chaos} 30, 051104.
\newblock \doi{https://doi.org/10.1063/5.0008341}
\input{DRA20}

\bibitem[{Eichinger and Lechner(2004)}]{EIC04}
Eichinger, S. and Lechner, K. (2004).
\newblock \emph{H{\"a}morrhagische und thromboembolische Komplikationen bei
  malignen Erkrankungen} (Springer), chap.~34.
\newblock 799--809.
\newblock \doi{https://doi.org/10.1007/978-3-662-06670-6_34}
\input{EIC04}

\bibitem[{Elisia et~al.(2017)Elisia, Lam, Hofs, Li, Hay, Cho et~al.}]{ELI17}
Elisia, I., Lam, V., Hofs, E., Li, M.~Y., Hay, M., Cho, B., et~al. (2017).
\newblock Effect of age on chronic inflammation and responsiveness to bacterial
  and viral challenges.
\newblock \emph{PLoS One} 12, e0188881.
\newblock \doi{https://doi.org/10.1371/journal.pone.0188881}
\input{ELI17}

\bibitem[{Fasshauer et~al.(2004)Fasshauer, Klein, Bl{\"u}her, and
  Paschke}]{FAS04}
Fasshauer, M., Klein, J., Bl{\"u}her, M., and Paschke, R. (2004).
\newblock Adipokine: M{\"o}gliches bindeglied zwischen insulinresistenz und
  adipositas.
\newblock \emph{Dtsch. {\"A}rztebl. Int.} 101, 3491
\input{FAS04}

\bibitem[{Franceschi and Campisi(2014)}]{FRA14}
Franceschi, C. and Campisi, J. (2014).
\newblock Chronic inflammation (inflammaging) and its potential contribution to
  age-associated diseases.
\newblock \emph{J. Gerontol. A Biol. Sci. Med. Sci.} 69, 4.
\newblock \doi{doi:10.1093/gerona/glu057}
\input{FRA14}

\bibitem[{Franceschi et~al.(2018)Franceschi}]{FRA18a}
Franceschi, C., Garagnani, P., Parini, P., Giuliani, C., Santoro, A. (2018).
\newblock Inflammaging: a new immune–metabolic viewpoint for age-related diseases.
\newblock \emph{Nature Reviews Endocrinology} 14, 576.
\input{FRA18a}

\bibitem[{Fulop et~al.(2018)Fulop, Larbi, Dupuis, Le~Page, Frost, Cohen
  et~al.}]{FUL18}
Fulop, T., Larbi, A., Dupuis, G., Le~Page, A., Frost, E.~H., Cohen, A.~A.,
  et~al. (2018).
\newblock immunosenescence and inflamm-aging as two sides of the same coin:
  Friends or foes?
\newblock \emph{Front. Immunol.} 8, 1960.
\newblock \doi{10.3389/fimmu.2017.01960}
\input{FUL18}

\bibitem[{Gaillard(2007)}]{GAI07}
Gaillard, R.~C. (2007).
\newblock Adipozyten: endokrine hochleistungsfabriken.
\newblock \emph{Cardiovasc. Med.} 10, 163.
\newblock \doi{https://doi.org/10.4414/cvm.2007.01248}
\input{GAI07}

\bibitem[{Gastpar(1982)}]{GAS82}
Gastpar, H. (1982).
\newblock \emph{Die Beeinflussung der Metastasierung {\"u}ber
  Gerinnungsmechanismen} (Thieme), chap. unb.
\newblock 118--128
\input{GAS82}

\bibitem[{Greten and Grivennikov(2019)}]{GRE19}
Greten, F.~R. and Grivennikov, S.~I. (2019).
\newblock {I}nflammation and {C}ancer: {T}riggers, {M}echanisms, and
  {C}onsequences.
\newblock \emph{Immunity} 51, 27.
\newblock \doi{https://doi.org/10.1016/j.immuni.2019.06.025}
\input{GRE19}

\bibitem[{Gross and Blasius(2008)}]{GRO08a}
Gross, T. and Blasius, B. (2008).
\newblock Adaptive coevolutionary networks: a review.
\newblock \emph{J. R. Soc. Interface} 5, 259--271.
\newblock \doi{10.1098/rsif.2007.1229}
\input{GRO08a}

\bibitem[{Gross et~al.(2006)Gross, D'Lima, and Blasius}]{GRO06b}
Gross, T., D'Lima, C. J.~D., and Blasius, B. (2006).
\newblock Epidemic dynamics on an adaptive network.
\newblock \emph{Phys. Rev. Lett.} 96, 208701.
\newblock \doi{10.1103/physrevlett.96.208701}
\input{GRO06b}

\bibitem[{Heerboth et~al.(2015)Heerboth, Housman, Leary, Longacre, Byler,
  Lapinska et~al.}]{HEE15a}
Heerboth, S., Housman, G., Leary, M., Longacre, M., Byler, S., Lapinska, K.,
  et~al. (2015).
\newblock {EMT} and tumor metastasis.
\newblock \emph{Clin. Transl. Med.} 4, 6.
\newblock \doi{https://doi.org/10.1186/s40169-015-0048-3}
\input{HEE15a}

\bibitem[{Hoppensteadt and Izhikevich(1996)}]{HOP96}
Hoppensteadt, F.~C. and Izhikevich, E.~M. (1996).
\newblock Synaptic organizations and dynamical properties of weakly connected
  neural oscillators ii. learning phase information.
\newblock \emph{Biol. Cybern.} 75, 129 --135.
\newblock \doi{10.1007/s004220050280}
\input{HOP96}

\bibitem[{Hotchkiss and Moldawer(2014)}]{HOT14}
Hotchkiss, R.~S. and Moldawer, L.~L. (2014).
\newblock Parallels between cancer and infectious disease.
\newblock \emph{N. Engl. J. Med.} 371, 380--383.
\newblock \doi{10.1056/nejmcibr1404664}
\input{HOT14}

\bibitem[{Hotchkiss et~al.(2016)Hotchkiss, Moldawer, Opal, Reinhart, Turnbull,
  and Vincent}]{HOT16}
Hotchkiss, R.~S., Moldawer, L.~L., Opal, S.~M., Reinhart, K., Turnbull, I.~R.,
  and Vincent, J.~L. (2016).
\newblock Sepsis and septic shock.
\newblock \emph{Nat. Rev. Dis. Primers} 2, 16045.
\newblock \doi{10.1038/nrdp.2016.45}
\input{HOT16}

\bibitem[{Majetschak et~al.(2001)Majetschak, Schade}]{MAJ01}
Majetschak, M., Schade, V.~A. (2001).
\newblock Mechanismen der inflammatorischen Wirtsantwort bei schweren Infektionen.
\newblock Intensivmedizin; Thieme Stuttgart, New York.
\input{MAJ01}

\bibitem[{Thomas(2020)}]{THO20}
Thomas, L. (2020).
\newblock Labor und Diagnose.
\newblock https://www.clinical-laboratory-diagnostics-2020.com/.
\input{THO20}

\bibitem[{Ivanov and Bartsch(2014)}]{IVA14}
Ivanov, P.~C. and Bartsch, R.~P. (2014).
\newblock \emph{Network {P}hysiology: {M}apping {I}nteractions {B}etween
  {N}etworks of {P}hysiologic {N}etworks} (Springer), chap.~10.
\newblock Networks of {N}etworks: {T}he {L}ast {F}rontier of {C}omplexity.
  203--222.
\newblock \doi{978-3-319-37492-5}
\input{IVA14}

\bibitem[{Ivanov et~al.(2009)Ivanov, Ma, and Bartsch}]{IVA09}
Ivanov, P.~C., Ma, Q. D.~Y., and Bartsch, R.~P. (2009).
\newblock Maternal - fetal heartbeat phase synchronization.
\newblock \emph{Proc. Natl. Acad. Sci. U.S.A.} 106, 13641.
\newblock \doi{https://doi.org/10.1073/pnas.0906987106}
\input{IVA09}

\bibitem[{Jain and Krishna(2001)}]{JAI01}
Jain, S. and Krishna, S. (2001).
\newblock A model for the emergence of cooperation, interdependence, and
  structure in evolving networks.
\newblock \emph{Proc. Natl. Acad. Sci.} 98, 543--547.
\newblock \doi{10.1073/pnas.98.2.543}
\input{JAI01}

\bibitem[{Junqueira et~al.(1995)Junqueira, Carneiro, and Kelley}]{JUN95}
Junqueira, L.~C., Carneiro, J., and Kelley, R.~O. (1995).
\newblock \emph{Basic Histology} (New York City: McGraw Hill Education)
\input{JUN95}

\bibitem[{Karlsson et~al.(2017)Karlsson, Gonzalez, Welin, and Fuxe}]{KAR17a}
Karlsson, M.~C., Gonzalez, S.~F., Welin, J., and Fuxe, J. (2017).
\newblock Epithelial-mesenchymal transition in cancer metastasis through the
  lymphatic system.
\newblock \emph{Mol. Oncol.} 11, 781.
\newblock \doi{10.1002/1878-0261.12092}
\input{KAR17a}

\bibitem[{Kasatkin and Nekorkin(2018)}]{KAS18}
Kasatkin, D.~V. and Nekorkin, V.~I. (2018).
\newblock Synchronization of chimera states in a multiplex system of phase
  oscillators with adaptive couplings.
\newblock \emph{Chaos} 28, 093115.
\newblock \doi{10.1063/1.5031681}
\input{KAS18}

\bibitem[{Kasatkin et~al.(2017)Kasatkin, Yanchuk, Sch{\"o}ll, and
  Nekorkin}]{KAS17}
Kasatkin, D.~V., Yanchuk, S., Sch{\"o}ll, E., and Nekorkin, V.~I. (2017).
\newblock {S}elf-organized emergence of multi-layer structure and chimera
  states in dynamical networks with adaptive couplings.
\newblock \emph{Phys. Rev. E} 96, 062211.
\newblock \doi{10.1103/physreve.96.062211}
\input{KAS17}

\bibitem[{Kivel{\"a} et~al.(2014)Kivel{\"a}, Arenas, Barth{\'e}lemy, Gleeson,
  Moreno, and Porter}]{KIV14}
Kivel{\"a}, M., Arenas, A., Barth{\'e}lemy, M., Gleeson, J.~P., Moreno, Y., and
  Porter, M.~A. (2014).
\newblock Multilayer networks.
\newblock \emph{J. Complex Netw.} 2, 203--271.
\newblock \doi{10.1093/comnet/cnu016}
\input{KIV14}

\bibitem[{Kuehn(2015)}]{KUE15}
Kuehn, C. (2015).
\newblock \emph{Multiple Time Scale Dynamics} (Springer, Cham).
\newblock \doi{10.1007/978-3-319-12316-5}
\input{KUE15}

\bibitem[{Kuramoto and Battogtokh(2002)}]{KUR02a}
Kuramoto, Y. and Battogtokh, D. (2002).
\newblock {Coexistence of Coherence and Incoherence in Nonlocally Coupled Phase
  Oscillators.}
\newblock \emph{Nonlin. Phen. in Complex Sys.} 5, 380--385
\input{KUR02a}

\bibitem[{Lamouille et~al.(2014)Lamouille, Xu, and Derynck}]{LAM14a}
Lamouille, S., Xu, J., and Derynck, R. (2014).
\newblock Molecular mechanisms of epithelial-mesenchymal transition.
\newblock \emph{Nat. Rev. Mol. Cell Biol.} 15, 178.
\newblock \doi{https://doi.org/10.1038/nrm3758}
\input{LAM14a}

\bibitem[{Leyva et~al.(2017)Leyva, Sevilla-Escoboza, Sendi{\~n}a-Nadal,
  Guti{\'e}rrez, Buld{\'u}, and Boccaletti}]{LEY17a}
Leyva, I., Sevilla-Escoboza, R., Sendi{\~n}a-Nadal, I., Guti{\'e}rrez, R.,
  Buld{\'u}, J.~M., and Boccaletti, S. (2017).
\newblock Inter-layer synchronization in non-identical multi-layer networks.
\newblock \emph{Sci. Rep.} 7, 45475.
\newblock \doi{10.1038/srep45475}
\input{LEY17a}

\bibitem[{Lin et~al.(2016)Lin, Liu, Bartsch, and Ivanov}]{LIN16d}
Lin, A., Liu, K. K.~L., Bartsch, R.~P., and Ivanov, P.~C. (2016).
\newblock Delay-correlation landscape reveals characteristic time delays of
  brain rhythms and heart interactions.
\newblock \emph{Phil. Trans. R. Soc. A} 374, 20150182.
\newblock \doi{10.1098/rsta.2015.0182}
\input{LIN16d}

\bibitem[{Lippman(2016)}]{LIP16}
Lippman, M.~E. (2016).
\newblock \emph{Mammakarzinom} (ABW Wissenschaftsverlag), chap.~39.
\newblock 633--649
\input{LIP16}

\bibitem[{Liu et~al.(2013{\natexlab{a}})Liu, Aihara, and Chen}]{LIU13b}
Liu, R., Aihara, K., and Chen, L. (2013{\natexlab{a}}).
\newblock Dynamical network biomarkers for identifying critical transitions and
  their driving networks of biologic processes.
\newblock \emph{Quantitative Biology} 1, 105--114
\input{LIU13b}

\bibitem[{Liu et~al.(2012)Liu, Li, Liu, Wu, Chen, and Aihara}]{LIU12b}
Liu, R., Li, M., Liu, Z.~P., Wu, J., Chen, L., and Aihara, K. (2012).
\newblock Identifying critical transitions and their leading biomolecular
  networks in complex diseases.
\newblock \emph{Sci. Rep.} 2, 813
\input{LIU12b}

\bibitem[{Liu et~al.(2013{\natexlab{b}})Liu, Wang, Aihara, and Chen}]{LIU13a}
Liu, R., Wang, X., Aihara, K., and Chen, L. (2013{\natexlab{b}}).
\newblock {{E}arly diagnosis of complex diseases by molecular biomarkers,
  network biomarkers, and dynamical betwork biomarkers}.
\newblock \emph{Med. Res. Rev.}
\input{LIU13a}

\bibitem[{Longo(2011)}]{LON11}
Longo, D.~L. (2011).
\newblock \emph{Harrison's Hematology and Oncology} (New York City: McGraw-Hill
  Companies)
\input{LON11}

\bibitem[{L{\"o}ser(2018)}]{LOE18a}
L{\"o}ser, T. (2018).
\newblock Process analysis of carcinogenesis: concept derivation of the tissue
  function "preservation of a homogeneous gene expression".
\newblock \emph{Theory Biosci.} 137, 85.
\newblock \doi{https://doi.org/10.1007/s12064-017-0256-z}
\input{LOE18a}

\bibitem[{L{\"o}ser(2020)}]{LOE20}
L{\"o}ser, T. (2020).
\newblock Aspects of tumor progression.
\newblock \emph{Med. Hypotheses} 144, 110157.
\newblock \doi{https://doi.org/10.1016/j.mehy.2020.110157}
\input{LOE20}

\bibitem[{L\"ucken et~al.(2016)L\"ucken, Popovych, Tass, and Yanchuk}]{LUE16}
L\"ucken, L., Popovych, O.~V., Tass, P.~A., and Yanchuk, S. (2016).
\newblock {N}oise-enhanced coupling between two oscillators with long-term
  plasticity.
\newblock \emph{Phys. Rev. E} 93, 032210.
\newblock \doi{10.1103/physreve.93.032210}
\input{LUE16}

\bibitem[{Madadi~Asl et~al.(2018)Madadi~Asl, Valizadeh, and Tass}]{ASL18a}
Madadi~Asl, M., Valizadeh, A., and Tass, P.~A. (2018).
\newblock Dendritic and axonal propagation delays may shape neuronal networks
  with plastic synapses.
\newblock \emph{Front. Physiol.} 9, 1849.
\newblock \doi{10.3389/fphys.2018.01849}
\input{ASL18a}

\bibitem[{Maistrenko et~al.(2007)Maistrenko, Lysyansky, Hauptmann, Burylko, and
  Tass}]{MAI07}
Maistrenko, Y., Lysyansky, B., Hauptmann, C., Burylko, O., and Tass, P.~A.
  (2007).
\newblock Multistability in the kuramoto model with synaptic plasticity.
\newblock \emph{Phys. Rev. E} 75, 066207.
\newblock \doi{10.1103/physreve.75.066207}
\input{MAI07}

\bibitem[{Maksimenko et~al.(2016)Maksimenko, Makarov, Bera, Ghosh, Dana,
  Goremyko et~al.}]{MAK16}
Maksimenko, V.~A., Makarov, V.~V., Bera, B.~K., Ghosh, D., Dana, S.~K.,
  Goremyko, M.~V., et~al. (2016).
\newblock Excitation and suppression of chimera states by multiplexing.
\newblock \emph{Phys. Rev. E} 94, 052205.
\newblock \doi{10.1103/physreve.94.052205}
\input{MAK16}

\bibitem[{Male et~al.(2012)Male, Brostoff, Roth, and Roitt}]{MAL12}
Male, D., Brostoff, J., Roth, D., and Roitt, I. (2012).
\newblock \emph{Immunology} (Philadelphia: Saunders), 8th edn.
\input{MAL12}

\bibitem[{Mantovani et~al.(2017)Mantovani, Marchesi, Malesci, Laghi, and
  Allavena}]{MAN17}
Mantovani, A., Marchesi, F., Malesci, A., Laghi, L., and Allavena, P. (2017).
\newblock Tumour-associated macrophages as treatment targets in oncology.
\newblock \emph{Nat. Rev. Clin. Oncol.} 14, 399.
\newblock \doi{10.1038/nrclinonc.2016.217}
\input{MAN17}

\bibitem[{Markram et~al.(1997)Markram, L\"ubke, Frotscher, and
  Sakmann}]{MAR97a}
Markram, H., L\"ubke, J., Frotscher, M., and Sakmann, B. (1997).
\newblock Regulation of synaptic efficacy by coincidence of postsynaptic {AP}s
  and {EPSP}s.
\newblock \emph{Science} 275, 213--215.
\newblock \doi{10.1126/science.275.5297.213}
\input{MAR97a}

\bibitem[{Matsumoto et~al.(2018)Matsumoto, Ogura, Shimizu, Ikeda, Hirose,
  Matsuura et~al.}]{MAT18c}
Matsumoto, H., Ogura, H., Shimizu, K., Ikeda, M., Hirose, T., Matsuura, H.,
  et~al. (2018).
\newblock The clinical importance of a cytokine network in the acute phase of
  sepsis.
\newblock \emph{Sci. Rep.} 8, 13995.
\newblock \doi{https://doi.org/10.1038/s41598-018-32275-8}
\input{MAT18c}

\bibitem[{Meisel and Gross(2009)}]{MEI09a}
Meisel, C. and Gross, T. (2009).
\newblock Adaptive self-organization in a realistic neural network model.
\newblock \emph{Phys. Rev. E} 80, 061917.
\newblock \doi{10.1103/physreve.80.061917}
\input{MEI09a}

\bibitem[{Moorman et~al.(2016)Moorman, Lake, and Ivanov}]{MOO16}
Moorman, J.~R., Lake, D.~E., and Ivanov, P.~C. (2016).
\newblock Early {D}etection of {S}epsis - {A} {R}ole for {N}etwork
  {P}hysiology?
\newblock \emph{Crit. Care Med.} 44, 312.
\newblock \doi{10.1097/ccm.0000000000001548}
\input{MOO16}

\bibitem[{Mor{\'a}n et~al.(2013)Mor{\'a}n, Parra-Medina, Cardona,
  Quintero-Ronderos, and Rodr{\'i}guez}]{MOR13a}
Mor{\'a}n, G. A.~G., Parra-Medina, R., Cardona, A.~G., Quintero-Ronderos, P.,
  and Rodr{\'i}guez, {\'E}.~G. (2013).
\newblock Cytokines, chemokines and growth factors.
\newblock In \emph{Autoimmunity: From Bench to Bedside}, eds. J.~M. Anaya,
  Y.~Shoenfeld, A.~Rojas-Villarraga, R.~A. Levy, and R.~Cervera (Bogota,
  Colombia: El Rosario University Press), chap.~9. 133--168
\input{MOR13a}

\bibitem[{Nemetschek(1971)}]{NEM71}
Nemetschek, T. (1971).
\newblock \emph{Altersabh{\"a}ngige Abl{\"a}ufe am Kollagen in} (Schattauer
  Verlag), vol.~3 of \emph{Altern und Entwicklung}, chap.~3.
\newblock 38--68
\input{NEM71}

\bibitem[{Newman(2003)}]{NEW03}
Newman, M. E.~J. (2003).
\newblock The structure and function of complex networks.
\newblock \emph{SIAM Review} 45, 167--256.
\newblock \doi{10.1137/s0036144503}
\input{NEW03}

\bibitem[{Nikitin et~al.(2019)Nikitin, Omelchenko, Zakharova, Avetyan, Fradkov,
  and Sch{\"o}ll}]{NIK19}
Nikitin, D., Omelchenko, I., Zakharova, A., Avetyan, M., Fradkov, A.~L., and
  Sch{\"o}ll, E. (2019).
\newblock {C}omplex partial synchronization patterns in networks of
  delay-coupled neurons.
\newblock \emph{Phil. Trans. R. Soc. A} 377, 20180128.
\newblock \doi{10.1098/rsta.2018.0128}
\input{NIK19}

\bibitem[{Omelchenko et~al.(2019)Omelchenko, H{\"u}lser, Zakharova, and
  Sch{\"o}ll}]{OME19}
Omelchenko, I., H{\"u}lser, T., Zakharova, A., and Sch{\"o}ll, E. (2019).
\newblock Control of chimera states in multilayer networks.
\newblock \emph{Front. Appl. Math. Stat.} 4, 67.
\newblock \doi{10.3389/fams.2018.00067}
\input{OME19}

\bibitem[{Panaggio and Abrams(2015)}]{PAN15}
Panaggio, M.~J. and Abrams, D.~M. (2015).
\newblock Chimera states: Coexistence of coherence and incoherence in networks
  of coupled oscillators.
\newblock \emph{Nonlinearity} 28, R67.
\newblock \doi{10.1088/0951-7715/28/3/r67}
\input{PAN15}

\bibitem[{Pikovsky et~al.(2001)Pikovsky, Rosenblum, and Kurths}]{PIK01}
Pikovsky, A., Rosenblum, M., and Kurths, J. (2001).
\newblock \emph{Synchronization: a universal concept in nonlinear sciences}
  (Cambridge: Cambridge University Press), 1st edn.
\input{PIK01}

\bibitem[{Popovych et~al.(2013)Popovych, Yanchuk, and Tass}]{POP13}
Popovych, O.~V., Yanchuk, S., and Tass, P.~A. (2013).
\newblock Self-organized noise resistance of oscillatory neural networks with
  spike timing-dependent plasticity.
\newblock \emph{Sci. Rep.} 3, 2926.
\newblock \doi{10.1038/srep02926}
\input{POP13}

\bibitem[{Porporato(2016)}]{POR16b}
Porporato, P.~E. (2016).
\newblock Understanding cachexia as a cancer metabolism syndrome.
\newblock \emph{Oncogenesis} 5, pagee200.
\newblock \doi{https://doi.org/10.1038/oncsis.2016.3}
\input{POR16b}

\bibitem[{Prescott et~al.(2016)Prescott, Osterholzer, Langa, Angus, and
  Iwashyna}]{PRE16a}
Prescott, H.~C., Osterholzer, J.~J., Langa, K.~M., Angus, D.~C., and Iwashyna,
  T.~J. (2016).
\newblock Late mortality after sepsis: propensity matched cohort study.
\newblock \emph{BMJ} 353, i2375.
\newblock \doi{10.1136/bmj.i2375}
\input{PRE16a}

\bibitem[{Razak et~al.(2018)Razak, Jones, Bhandari, Berndt, and
  Metharon}]{RAZ18}
Razak, N. B.~A., Jones, G., Bhandari, M., Berndt, M.~C., and Metharon, P.
  (2018).
\newblock Cancer-associated thrombosis: An overview of mechanisms, risk
  factors, and treatment.
\newblock \emph{Cancers} 10, 380.
\newblock \doi{https://doi.org/10.3390/cancers10100380}
\input{RAZ18}

\bibitem[{Rich and Chaplin(2019)}]{RIC19}
Rich, R.~R. and Chaplin, D.~D. (2019).
\newblock \emph{The human immune response} (Amsterdam: Elsevier), chap.~1.
\newblock 3--17.
\newblock \doi{https://doi.org/10.1016/b978-0-7020-6896-6.00001-6}
\input{RIC19}

\bibitem[{R{\"o}hr et~al.(2019)R{\"o}hr, Berner, Lameu, Popovych, and
  Yanchuk}]{ROE19a}
R{\"o}hr, V., Berner, R., Lameu, E.~L., Popovych, O.~V., and Yanchuk, S.
  (2019).
\newblock Frequency cluster formation and slow oscillations in neural
  populations with plasticity.
\newblock \emph{PLoS ONE} 14, e0225094.
\newblock \doi{10.1371/journal.pone.0225094}
\input{ROE19a}

\bibitem[{Rybalova et~al.(2019)Rybalova, Vadivasova, Strelkova, Anishchenko,
  and Zakharova}]{RYB19}
Rybalova, E., Vadivasova, T., Strelkova, G., Anishchenko, V., and Zakharova, A.
  (2019).
\newblock Forced synchronization of a multilayer heterogeneous network of
  chaotic maps in the chimera state mode.
\newblock \emph{Chaos} 29, 033134.
\newblock \doi{10.1063/1.5090184}
\input{RYB19}

\bibitem[{Sakaguchi and Kuramoto(1986)}]{SAK86}
Sakaguchi, H. and Kuramoto, Y. (1986).
\newblock A soluble active rotater model showing phase transitions via mutual
  entertainment.
\newblock \emph{Prog. Theor. Phys} 76, 576--581
\input{SAK86}

\bibitem[{Sawicki(2019)}]{SAW20}
Sawicki, J. (2019).
\newblock \emph{Delay controlled partial synchronization in complex networks}.
\newblock Springer Theses (Heidelberg: Springer).
\newblock \doi{10.1007/978-3-030-34076-6_5}
\input{SAW20}

\bibitem[{Sawicki et~al.(2021)Sawicki, Koulen, and Sch{\"o}ll}]{SAW21}
Sawicki, J., Koulen, J.~M., and Sch{\"o}ll, E. (2021).
\newblock Synchronization scenarios in three-layer networks with a hub.
\newblock \emph{Chaos} 31, 073131
\input{SAW21}

\bibitem[{Sawicki et~al.(2018)Sawicki, Omelchenko, Zakharova, and
  Sch{\"o}ll}]{SAW18c}
Sawicki, J., Omelchenko, I., Zakharova, A., and Sch{\"o}ll, E. (2018).
\newblock {D}elay controls chimera relay synchronization in multiplex networks.
\newblock \emph{Phys. Rev. E} 98, 062224.
\newblock \doi{10.1103/physreve.98.062224}
\input{SAW18c}

\bibitem[{Sch{\"o}ll(2020)}]{SCH20}
Sch{\"o}ll, E. (2020).
\newblock Chimeras in physics and biology: Synchronization and
  desynchronization of rhythms.
\newblock \emph{Nova Acta Leopoldina} 425, 67--95.
\newblock Invited contribution
\input{SCH20}

\bibitem[{Sch{\"o}ll et~al.(2020)Sch{\"o}ll, Zakharova, and Andrzejak}]{SCH20b}
Sch{\"o}ll, E., Zakharova, A., and Andrzejak, R.~G. (2020).
\newblock \emph{Chimera States in Complex Networks}.
\newblock Research Topic, Front. Appl. Math. Stat. (Lausanne: Frontiers Media
  SA).
\newblock \doi{10.3389/978-2-88963-311-1}.
\newblock Ebook
\input{SCH20b}

\bibitem[{Seliger et~al.(2002)Seliger, Young, and Tsimring}]{SEL02}
Seliger, P., Young, S.~C., and Tsimring, L.~S. (2002).
\newblock Plasticity and learning in a network of coupled phase oscillators.
\newblock \emph{Phys. Rev. E} 65, 041906.
\newblock \doi{10.1103/physreve.65.041906}
\input{SEL02}

\bibitem[{Seymour et~al.(2019)Seymour, Kennedy, Wang, Chang, Elliott, Xu
  et~al.}]{SEY19}
Seymour, C.~W., Kennedy, J.~N., Wang, S., Chang, C.~H., Elliott, C.~F., Xu, Z.,
  et~al. (2019).
\newblock Derivation, validation, and potential treatment implications of novel
  clinical phenotypes for sepsis.
\newblock \emph{JAMA} 321, 20.
\newblock \doi{10.1001/jama.2019.5791}
\input{SEY19}

\bibitem[{Shannon(1948)}]{SHA48}
Shannon, C.~E. (1948).
\newblock A mathematical theory of communication.
\newblock \emph{Bell Syst. Tech. J.} 27, 379.
\newblock \doi{https://doi.org/10.1002/j.1538-7305.1948.tb01338.x}
\input{SHA48}

\bibitem[{Shepelev et~al.(2021)Shepelev, Muni, Sch{\"o}ll, and
  Strelkova}]{SHE21}
Shepelev, I.~A., Muni, S.~S., Sch{\"o}ll, E., and Strelkova, G.~I. (2021).
\newblock Repulsive inter-layer coupling induces anti-phase synchronization.
\newblock \emph{Chaos} 31, 063116
\input{SHE21}

\bibitem[{Singer et~al.(2016)Singer, Deutschman, Seymour, Shankar-Hari, Annane,
  Bauer et~al.}]{SIN16b}
Singer, M., Deutschman, C.~S., Seymour, C.~W., Shankar-Hari, M., Annane, D.,
  Bauer, M., et~al. (2016).
\newblock The third international consensus definitions for sepsis and septic
  shock (sepsis-3).
\newblock \emph{JAMA} 315, 801.
\newblock \doi{10.1001/jama.2016.0287}
\input{SIN16b}

\bibitem[{Strogatz(2001)}]{STR01a}
Strogatz, S.~H. (2001).
\newblock Exploring complex networks.
\newblock \emph{Nature} 410, 268--276.
\newblock \doi{10.1038/35065725}
\input{STR01a}

\bibitem[{Thomas(1972)}]{THO72}
Thomas, L. (1972).
\newblock Germs.
\newblock \emph{N. Engl. J. Med.} 287, 553.
\newblock \doi{10.1056/nejm197209142871109}
\input{THO72}

\bibitem[{Thomas(2020)}]{THO20}
Thomas, L. (2020).
\newblock Clinical Laboratory Diagnostics.
\newblock https://www.clinical-laboratory-diagnostics-2020.com/.
\input{THO20}

\bibitem[{Timms and English(2014)}]{TIM14}
Timms, L. and English, L.~Q. (2014).
\newblock {S}ynchronization in phase-coupled {K}uramoto oscillator networks
  with axonal delay and synaptic plasticity.
\newblock \emph{Phys. Rev. E} 89, 032906.
\newblock \doi{10.1103/physreve.89.032906}
\input{TIM14}

\bibitem[{Tragl(1999)}]{TRA99a}
Tragl, K.~H. (1999).
\newblock \emph{Handbuch der internistischen Geriatrie} (Springer)
\input{TRA99a}

\bibitem[{Vineis et~al.(2010)Vineis, Schatzkin, and Potter}]{VIN10}
Vineis, P., Schatzkin, A., and Potter, J.~D. (2010).
\newblock Models of carcinogenesis: an overview.
\newblock \emph{Carcinogenesis} 31, 1703.
\newblock \doi{https://doi.org/10.1093/carcin/bgq087}
\input{VIN10}

\bibitem[{Virchow(1978)}]{VIR78}
Virchow, R. (1978).
\newblock \emph{Die krankhaften Geschw{\"u}lste} (Springer)
\input{VIR78}

\bibitem[{Walther(1948)}]{SCH48}
Walther, H.~E. (1948).
\newblock \emph{Krebsmetastasen} (Benno Schwabe Verlag)
\input{SCH48}

\bibitem[{Warburg et~al.(1924)Warburg, Posener, and Negelein}]{WAR24}
Warburg, O., Posener, K., and Negelein, E. (1924).
\newblock \"{U}ber den {S}toffwechsel der {C}arcinomzelle.
\newblock \emph{Biochem. Z.} 152, 309
\input{WAR24}

\bibitem[{Weinberg(2014)}]{WEI14d}
Weinberg, R.~A. (2014).
\newblock \emph{The biology of cancer} (Milton: Garland Publishing Inc.), 2nd
  edn.
\input{WEI14d}

\bibitem[{Wu and Zhou(2009)}]{WU09a}
Wu, Y. and Zhou, B.~P. (2009).
\newblock Inflammation: a driving force speeds cancer metastasis.
\newblock \emph{Cell Cycle} 8, 3267.
\newblock \doi{10.4161/cc.8.20.9699}
\input{WU09a}

\bibitem[{Xu et~al.(2006)Xu, Chen, Hu, Stanley, and Ivanov}]{XU06a}
Xu, L., Chen, Z., Hu, K., Stanley, H.~E., and Ivanov, P.~C. (2006).
\newblock Spurious detection of phase synchronization in coupled nonlinear
  oscillators.
\newblock \emph{Phys. Rev. E} 73, 065201.
\newblock \doi{10.1103/physreve.73.065201}
\input{XU06a}

\bibitem[{Yiu et~al.(2012)Yiu, Graham, and Stengel}]{YIU12}
Yiu, H.~H., Graham, A.~L., and Stengel, R.~F. (2012).
\newblock Dynamics of a cytokine storm.
\newblock \emph{PLoS ONE} 7, 1--15.
\newblock \doi{10.1371/journal.pone.0045027}
\input{YIU12}

\bibitem[{Zhang et~al.(2015)Zhang, Boccaletti, Guan, and Liu}]{ZHA15a}
Zhang, X., Boccaletti, S., Guan, S., and Liu, Z. (2015).
\newblock Explosive synchronization in adaptive and multilayer networks.
\newblock \emph{Phys. Rev. Lett.} 114, 038701.
\newblock \doi{10.1103/physrevlett.114.038701}
\input{ZHA15a}

\bibitem[{Zhang et~al.(2021)Zhang, Zuo, Liu, Hu, Yang, Qiu et~al.}]{ZHA21d}
Zhang, Y., Zuo, C., Liu, L., Hu, Y., Yang, B., Qiu, S., et~al. (2021).
\newblock Single-cell {RNA}-sequencing atlas reveals an {MDK}-dependent
  immunosuppressive environment in {E}rb{B} pathway-mutated gallbladder cancer.
\newblock \emph{J. Hepatol.} ,
  1\doi{https://doi.org/10.1016/j.jhep.2021.06.023}
\input{ZHA21d}

\end{thebibliography}

\end{document}